\documentclass[twocolumn]{aastex631}
\usepackage{threeparttable}
\usepackage{mathrsfs}
\received{Mar 2 2022}
\revised{YYY}
\accepted{ZZZ}

\newcommand{\gsup}{3.04}
\newcommand{\gsuperr}{0.80}
\newcommand{\rsup}{36.6}
\newcommand{\rsuperr}{2.7}
\newcommand{\Menc}{9.26}
\newcommand{\dMenc}{2.78}

\newcommand{\fhostcore}{16.3}
\newcommand{\fhostnocore}{41.1}

\newcommand{\mfardal}{12.23}
\newcommand{\mfardalerr}{0.10}
\newcommand{\Mencfardalerr}{0.49} 
\newcommand{\Mencfardal}{3.48}
\newcommand{\Mencfardalbary}{4.55} 
\newcommand{\Mencoverestimate}{2.04} 
\newcommand{\dMencoverestimate}{2.82} 
\newcommand{\Mdiskatshell}{7.29}
\newcommand{\Mvirsatlo}{1.0}
\newcommand{\Mvirsathi}{1.8}

\newcommand{\fehphotmedne}{$-0.41$}
\newcommand{\fehphotmedall}{$-0.43$}
\newcommand{\fehphotmedmem}{$-0.42$}
\newcommand{\fehphotavgall}{$-0.54$}
\newcommand{\fehphotavgmem}{$-0.54$}
\newcommand{\fehphotavgne}{$-0.53$}
\newcommand{\fehphotavgdiffhaloage}{$0.04$}
\newcommand{\fehphotavgdiffgssage}{$0.09$}
\newcommand{\fehphotmeddiffhaloage}{$0.03$}
\newcommand{\fehphotmeddiffgssage}{$0.08$}
\newcommand{\fehphoterrmed}{$0.05$}

\newcommand{\fehphotmeddiffnenodisk}{0.04}
\newcommand{\fehphotavgdiffnenodisk}{0.04}
\newcommand{\fehphotmeddiffnenodisksix}{0.02}
\newcommand{\fehphotavgdiffnenodisksix}{0.04}
\newcommand{\fehphotmednepsub}{$-0.42$}
\newcommand{\fehphoterrmednepsub}{0.01}

\newcommand{\psubmaxgssne}{95\%}
\newcommand{\psubmaxse}{82\%}
\newcommand{\psubmaxw}{89\%}

\newcommand{\vdisknethree}{$-242.2$} 
\newcommand{\vdisknefour}{$-193.0$} 
\newcommand{\vdiskneone}{$-227.9$} 
\newcommand{\vdisknesix}{$-202.7$} 
\newcommand{\vdisknetwo}{$-206.6$} 
\newcommand{\alphagamma}{2.25}
\newcommand{\betagamma}{506.25}
\newcommand{\rgbad}{63}

\newcommand{\vsyssimongeha}{2.2}
\newcommand{\kms}{km s$^{-1}$}
\newcommand{\vtotmed}{5.3}
\newcommand{\vsys}{$-300$}

\newcommand{\fehcaterrmax}{0.8}
\newcommand{\ewnaerrmax}{2}

\newcommand{\muvmw}{$-$62.6}
\newcommand{\sigvmw}{39.6}
\newcommand{\percentmwvel}{82.8\%}

\newcommand{\Ntot}{556}
\newcommand{\medewnaand}{1.7}
\newcommand{\medewnamw}{3.1}
\newcommand{\medfehcatand}{$-0.54$}
\newcommand{\medfehcatmw}{$-1.74$}

\newcommand{\alphafe}{[$\alpha$/Fe]}
\newcommand{\feh}{[Fe/H]}
\newcommand{\fehphot}{[Fe/H]$_{\rm phot}$}
\newcommand{\vhelio}{$v_{\rm helio}$}
\newcommand{\fehcat}{[Fe/H]$_{\rm CaT}$}
\newcommand{\ewna}{EW$_{\rm Na}$}
\newcommand{\fehcaterr}{$\delta$\fehcat}
\newcommand{\ewnaerr}{$\delta$\ewna}
\newcommand{\Rproj}{R$_{\rm proj}$}
\newcommand{\Rdisk}{R$_{\rm disk}$}

\newcommand{\dm}{24.47}

\submitjournal{AJ}

\shorttitle{NE Shelf}
\shortauthors{Escala et al.}

\graphicspath{{./}{figures/}}

\begin{document}

\title{Kinematics and Metallicity of Red Giant Branch Stars in the Northeast Shelf of M31\footnote{The data presented herein were obtained at the W. M. Keck Observatory, which is operated as a scientific partnership among the California Institute of Technology, the University of California and the National Aeronautics and Space Administration. The Observatory was made possible by the generous financial support of the W. M. Keck Foundation.}}

\correspondingauthor{I. Escala}
\email{iescala@carnegiescience.edu}

\author[0000-0002-9933-9551]{Ivanna Escala}
\altaffiliation{Carnegie-Princeton Fellow}
\affiliation{The Observatories of the Carnegie Institution for Science, 813 Santa Barbara St, Pasadena, CA, 91101}
\affiliation{Department of Astrophysical Sciences, Princeton University, 4 Ivy Lane, Princeton, NJ, 08544, USA}

\author[0000-0003-0394-8377]{Karoline M. Gilbert}
\affiliation{Space Telescope Science Institute, 3700 San Martin Drive, Baltimore, MD 21218, USA}
\affiliation{The William H. Miller III Department of Physics \& Astronomy, Bloomberg Center for Physics and Astronomy, John Hopkins University, 3400 N. Charles St, Baltimore, MD 21218, USA}

\author[0000-0003-4207-3788]{Mark Fardal}
\affiliation{Space Telescope Science Institute, 3700 San Martin Drive, Baltimore, MD 21218, USA}

\author[0000-0001-8867-4234]{Puragra Guhathakurta}
\affiliation{UCO/Lick Observatory, Department of Astronomy \& Astrophysics, University of California Santa Cruz, 1156 High Street, Santa Cruz, California 95064, USA}

\author[0000-0003-3939-3297]{Robyn E. Sanderson}
\affiliation{Department of Physics \& Astronomy, University of Pennsylvania, 209 S 33rd Street, Philadelphia, PA 19104, USA}
\affiliation{Center for Computational Astrophysics, Flatiron Institute, 162 5th Avenue, New York, NY 10010, USA}

\author[0000-0001-9690-4159]{Jason S. Kalirai}
\affiliation{Johns Hopkins University Applied Physics Laboratory, 11100 Johns Hopkins Road, Laurel, MD 20723}

\author{Bahram Mobasher}
\affiliation{Department of Physics and Astronomy, University of California Riverside, 900 University Avenue, Riverside, CA 92521}

\begin{abstract}

We obtained Keck/DEIMOS spectra of \Ntot\ individual red giant branch stars in 4 spectroscopic fields spanning $13-31$ projected kpc along the Northeast (NE) shelf of M31. We present the first detection of a complete wedge pattern in the space of projected M31-centric radial distance versus line-of-sight velocity for this feature, which includes the returning stream component of the shelf. This wedge pattern agrees with expectations of a tidal shell formed in a radial merger and provides strong evidence in favor of predictions of Giant Stellar Stream (GSS) formation models in which the NE shelf originates from the second orbital wrap of the tidal debris.
The observed concentric wedge patterns of the NE, West (W), and Southeast (SE) shelves corroborates this interpretation independently of the models.
We do not detect a kinematical signature in the NE shelf region corresponding to an intact progenitor core, favoring GSS formation models in which the progenitor is completely disrupted. The shelf's photometric metallicity ([Fe/H]$_{\rm phot}$) distribution implies that it is dominated by tidal material, as opposed to the phase-mixed stellar halo or the disk. The metallicity distribution ([Fe/H]$_{\rm phot}$ = $-0.42$ $\pm$ $0.01$) also matches the GSS, and consequently the W and SE shelves, further supporting a direct physical association between the tidal features. 

\end{abstract}

\keywords{Galaxy structure (622) --- Andromeda galaxy (39) --- Galaxy mergers (608) --- Galaxy dynamics (591)}

\section{Introduction} \label{sec:intro}

In the hierarchical assembly paradigm, galaxy mergers play an essential role in driving the formation and growth of galaxies (e.g., \citealt{WhiteRees1978}). The tidal relics of these merger events encode a wealth of information concerning the accretion history of the host galaxy (e.g., \citealt{BullockJohnston2005,Johnston2008}), particularly in the form of stellar streams and shells  \citep{HernquistQuinn1988,Johnston2001,HendelJohnston2015}. The dynamics of streams and shells can also serve as powerful probes of the host galaxy gravitational potential \citep{MerrifieldKuijken1998,Ebrova2012,SandersonHelmi2013,LM10,Fardal2013}.

The Milky Way (MW) provides ample evidence for ongoing tidal disruption in the form of the Sagittarius stream \citep{Ibata2001sgr}, a plethora of smaller streams (\citealt{Helmi2020} and references therein), and potentially shell-like overdensities \citep{Belokurov2007,Juric2008} that may originate from a radial merger \citep{Simion2018,Simion2019,Donlon2019,Donlon2020,Naidu2021}. Deep, wide-field imaging surveys of massive galaxies beyond the Local Group have revealed rich collections of large-scale tidal features (including shells) surrounding both spirals and ellipticals (e.g., \citealt{Tal2009,MD10,MD21ssls,Atkinson2013,Kado-Fong2018}). However, since these low surface brightness features are intrinsically difficult to detect for large samples of external galaxies, many studies have focused on the resolved stellar populations of features in individual nearby galaxies (e.g., \citealt{Mouhcine2010,Ibata2014,Okamoto2015,Crnojevic2016,MD21}). Studies of kinematical tracers probing tidal debris are even more challenging and have thus been limited to a handful of external galaxies (e.g., \citealt{Romanowsky2012,Foster2014}).

Owing to its proximity ($d_{\rm M31}$ = 773 kpc; \citealt{Conn2016}), the Andromeda galaxy (M31) presents a rare opportunity for detailed photometric and spectroscopic studies of resolved stars in  tidal structures. M31 possesses a metal-rich giant stellar stream (GSS; \citealt{Ibata2001}) in the southern portion of its stellar halo and a diffuse shelf-like overdensity in the northeast (NE) with red colors similar to the GSS \citep{Ferguson2002}. \citet{Fardal2007} detected a second shelf in the west (W) using the \citeauthor{Ferguson2002} imaging and \citet{Gilbert2007} identified a third faint shelf in the southeast (SE) via spectroscopy. The chemical similarity of red giant branch (RGB) stars in the W and SE shelves to the GSS \citep{Gilbert2007,Fardal2012} as well as the striking consistency between various stellar populations in the NE, W, and SE shelves and the GSS \citep{Brown2006,Brown2008,Ferguson2005,Richardson2008,Tanaka2010,Bernard2015} support the hypothesis that these structures form a related system of tidal debris. The luminosity functions of planetary nebulae (PNe) in M31 additionally support a common origin for the NE shelf and GSS, while constraints for the W shelf are limited but suggest it is distinct from the disk \citep{Bhattacharya2019,Bhattacharya2021}. However, kinematical and chemical information in the NE shelf is necessary to finally confirm the putative association between the shelves and the GSS.

Orbital models for the formation of the GSS in a minor merger ($M_\star \sim (1-5) \times 10^9 M_\odot$) predict that the shelves correspond to tidal debris from the distinct progenitor of the GSS \citep{Fardal2006,Fardal2007,Fardal2008,Fardal2013,MoriRich2008,Sadoun2014,Miki2016,Kirihara2014,Kirihara2017}. In these models, the NE shelf is composed of material from the second pericentric passage of the disrupting progenitor as well as the forward continuation of the GSS. The models place the central material of the progenitor in the NE shelf region 
and allow for the presence of an intact GSS progenitor core based on its initial central density (e.g., \citealt{Fardal2013,Kirihara2017}), although no such structure has yet been detected (cf. \citealt{Dorman2012,Davidge2012}). Major merger models ($M_\star \sim 10^{10} M_\odot$) can similarly form tidal structures resembling the GSS and shelves \citep{Hammer2018,DSouzaBell2018}, but have not yet been shown to reproduce the shelves at the level of detail provided by minor merger models.

Despite not explicitly fitting for the detailed properties of the shelves, the \citeauthor{Fardal2013} minor merger models have predicted the existence of the SE shelf \citep{Fardal2007,Gilbert2007} and show excellent agreement with the kinematics of RGB stars in the W shelf \citep{Fardal2012}. These successes appear to be natural consequences of the constraints placed on the merger scenario using spatial and kinematical observations of the GSS \citep{McConnachie2003,Ibata2004,Gilbert2009} and stellar surface density maps of M31 \citep{Ibata2001,Ferguson2002,Irwin2005}, where the orbital trajectory of the progenitor is only loosely required to pass through the location of the NE shelf. Moreover, \citet{Fardal2013} demonstrated that the predicted kinematical signature for the NE shelf partially overlaps with a small sample of PNe that form an apparent stream near the shelf \citep{Merrett2003,Merrett2006}.

 Detailed kinematics for a large sample of RGB stars in the NE shelf could conclusively establish that it is indeed a tidal shell, place stringent constraints on GSS merger scenarios, identify whether an intact GSS progenitor core exists, and probe M31's gravitational potential at the location of the shelf. Furthermore, the NE shelf metallicity distribution could be used to infer the properties of the progenitor as the shelf region is predicted to contain its metal-rich central debris \citep{MoriRich2008,Fardal2008,Miki2016,Kirihara2017}. The observed metallicity gradient(s) in the GSS \citep{Ibata2007,Gilbert2009,Conn2016,Cohen2018,Escala2021} could be combined with the metallicity distributions of the NE, W, and SE shelves to connect chemical variations on the sky to the intrinsic properties of the progenitor \citep{Fardal2008,Miki2016,Kirihara2017,Milosevic2022}. This metallicity mapping may be crucial for distinguishing between major and minor merger scenarios for the formation of the GSS and shelves (see the discussions of \citealt{Gilbert2019} and \citealt{Escala2021}).

In this work, we present the first analysis of the metallicity and kinematics of RGB stars in the NE shelf of M31. This contribution is part of the Spectroscopic and Photometric Landscape of Andromeda's Stellar Halo survey (SPLASH; \citealt{Guhathakurta2005,Gilbert2006}), which has produced tens of thousands of Keck/DEIMOS spectra along lines-of-sight to M31's halo, disk, and dwarf galaxies (e.g., \citealt{Kalirai2010,Dorman2012,Dorman2015,Gilbert2012,Gilbert2014,Tollerud2012}). In Section~\ref{sec:data}, we introduce spectroscopic and photometric data for the NE shelf. We investigate the kinematics, metallicity, and projected phase space distribution of the NE shelf in Section~\ref{sec:props}. We perform detailed comparisons of the shelf's observed properties to N-body models in Section~\ref{sec:nbody} and the GSS and W and SE shelves as observed by SPLASH in Section~\ref{sec:w_se_gss}. Lastly, we discuss results for the NE shelf in the context of the literature in Section~\ref{sec:discuss} before summarizing in Section~\ref{sec:summary}.

\section{Data} \label{sec:data}

\subsection{Observations and Data Reduction} \label{sec:obs}

\begin{table*}
    \centering
    \begin{threeparttable}
    \caption{Keck/DEIMOS Multi-Object Slitmask Observations in M31's NE Shelf}
    \begin{tabular*}{\textwidth}{lcccccccccc}
        \hline \hline
        Slitmask & \multicolumn{1}{p{1.0cm}}{\centering R$_{\rm proj}$\\(kpc)} & R.A. & Decl. &  \multicolumn{1}{p{1.5cm}}{\centering Mask P.A.\\ ($^\circ$ E of N)} & \multicolumn{1}{p{1.5cm}}{\centering Obs. Date\\(UT)} &  \multicolumn{1}{p{1.3cm}}{\centering t$_{\rm exp}$\\(min)} & Airmass & \multicolumn{1}{p{1.2cm}}{\centering No.\\Targets\tnote{a}} & \multicolumn{1}{p{1.0cm}}{\centering No.\\RV\tnote{b}} & \multicolumn{1}{p{1.0cm}}{\centering No.\\M31\tnote{c}}\\ \hline
        NE1 & 19.6 & 00:50:04.96 & +41:39:53.3 & 30 & 2011 Nov 23 & 50.0 & 1.10 & 190 & 138 & 123\\
        NE2 & 25.7 & 00:51:52.11 & +42:02:22.3 & 30 & 2011 Nov 23 & 55.0 & 1.26 & 186 & 117 & 79 \\
        NE3 & 14.5 & 00:48:17.03 & +41:26:33.6 & 30 & 2011 Nov 24 & 51.7 & 1.27 & 214 & 154 & 135\\
        NE4 & 29.2 & 00:52:52.20 & +42:15:44.8 & 30 & 2011 Nov 24 & 60.0 & 1.08 & 182 & 124 & 91\\
        NE6 & 21.8 & 00:50:27.03 & +41:56:32.8 & 40 & 2011 Nov 24 & 51.7 & 1.23 & 200 & 149 & 128\\
        \hline
    \end{tabular*}
    \begin{tablenotes}
    \footnotesize
    \item[a] Number of targets on the slitmask with successfully extracted one-dimensional spectra.
    \item[b] Number of targets with successful radial velocity measurements (Section~\ref{sec:obs}).
    \item[c] Number of targets classified as M31 RGB stars (Section~\ref{sec:mem}).
    \end{tablenotes}
    \label{tab:obs}
    \end{threeparttable}
\end{table*}


\begin{figure}
    \centering
    \includegraphics[width=\columnwidth]{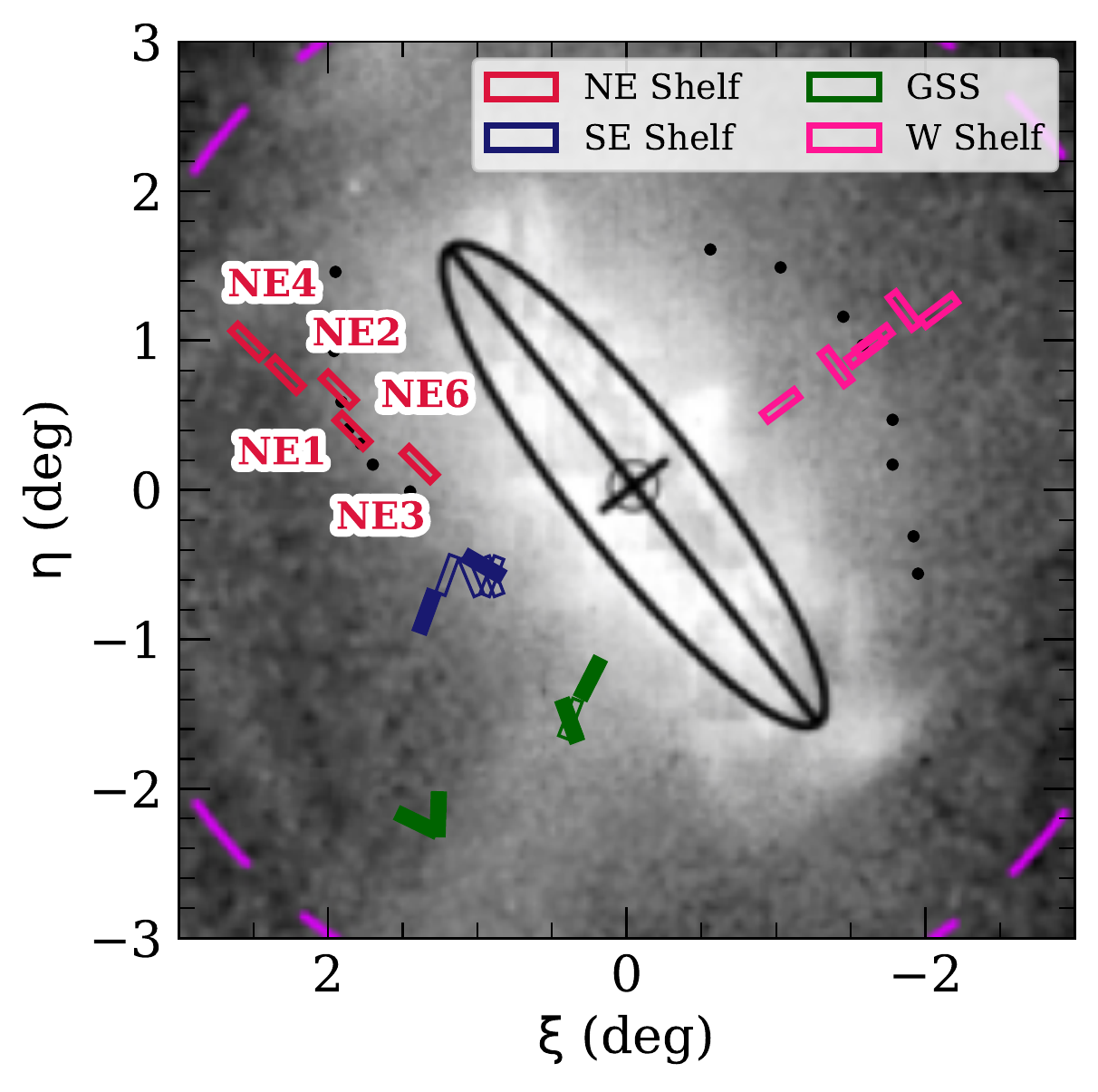}
    \caption{Location of the spectroscopic fields targeting the NE Shelf (red open boxes; Table~\ref{tab:obs}, Section~\ref{sec:obs}) in M31-centric coordinates overlaid on a star count map from the PAndAS survey \citep{McConnachie2018}. The approximate size and orientation of each DEIMOS field is represented by a 16' $\times$ 4' rectangle. We also show spectroscopic fields from the Elemental Abundances in M31 survey (filled boxes; \citealt{Gilbert2019,Escala2020a,Escala2020b}) and fields previously observed as part of the SPLASH survey (open boxes; \citealt{Kalirai2006,Gilbert2007,Gilbert2009,Fardal2012}) spanning the SE Shelf (blue), W shelf (pink), and GSS (green). 
    The black points are the edges of the NE and W shelves defined by \citet{Fardal2007} via applying an edge filter to the \citet{Irwin2005} star count map.}
    \label{fig:fields}
\end{figure}

We observed five slitmasks targeting M31's NE shelf with the DEIMOS instrument on Keck II as part of the SPLASH survey. We utilized the 1200 $\ell$/mm grating, which has a spectral dispersion of 0.33 \AA\ pixel$^{-1}$, with the OG550 order blocking filter and a central wavelength of 7800 \AA\ for our science  configuration. The spectroscopic fields were observed in Fall 2011 with $\sim$1 hr exposures. Table~\ref{tab:obs} summarizes the observations. Figure~\ref{fig:fields} shows the locations of the NE shelf fields on the sky in M31-centric coordinates, overlaid on a star count map of RGB candidates from the Pan-Andromeda Archaeological Survey (PAndAS; \citealt{McConnachie2018}). The naming convention for the NE shelf fields indicates the order in which the masks were designed, rather than their radial distance from M31. We included reference DEIMOS fields from the Elemental Abudances in M31 survey \citep{Gilbert2019,Escala2020a,Escala2020b} and fields previously observed as part of the SPLASH survey \citep{Kalirai2006,Gilbert2007,Gilbert2009,Fardal2012} targeting the SE shelf, W shelf, and GSS. Figure~\ref{fig:fields} also indicates the edges of the NE and W shelves defined by \citet{Fardal2007} via applying the Sobel operator to the \citet{Irwin2005} star map.

We extracted and reduced one-dimensional spectra from the two-dimensional images using a modified version of the {\tt\string spec2d} software \citep{Cooper2012,Newman2013} for stellar sources \citep{SimonGeha2007}. The one-dimensional spectra were then cross-correlated against empirical templates of hot stars to shift them into the rest frame and measure radial velocities. We corrected the velocity measurements for slit miscentering using the A-band \citep{Sohn2007} and transformed them into the heliocentric frame. As described by \citet{Howley2013}, the total velocity uncertainty is determined by scaling the cross-correlation based error with a value determined from duplicate measurements of RGB stars at the distance of M31, and then adding in quadrature with a systematic velocity error of \vsyssimongeha\ \kms\ computed by \citet{SimonGeha2007}. 

Using the visual inspection software {\tt\string zspec} (D.~Madgwick, DEEP2 survey), each spectrum was assigned a quality code ($Q$) for its radial velocity measurement. We restrict our analysis to objects with successful radial velocity measurements that match two or more strong spectral features ($Q = 4$) or a combination of one strong spectral feature and additional marginal features ($Q = 3$; \citealt{Guhathakurta2006}). The median total velocity uncertainty for stars with successful velocity measurements is \vtotmed\ \kms. We excluded some targets with successful velocity measurements from the subsequent analysis because of missing spectral data at red wavelengths, which is necessary to evaluate membership (Section~\ref{sec:mem}).

\subsection{Photometry}
\label{sec:phot}

\begin{figure}
    \centering
    \includegraphics[width=\columnwidth]{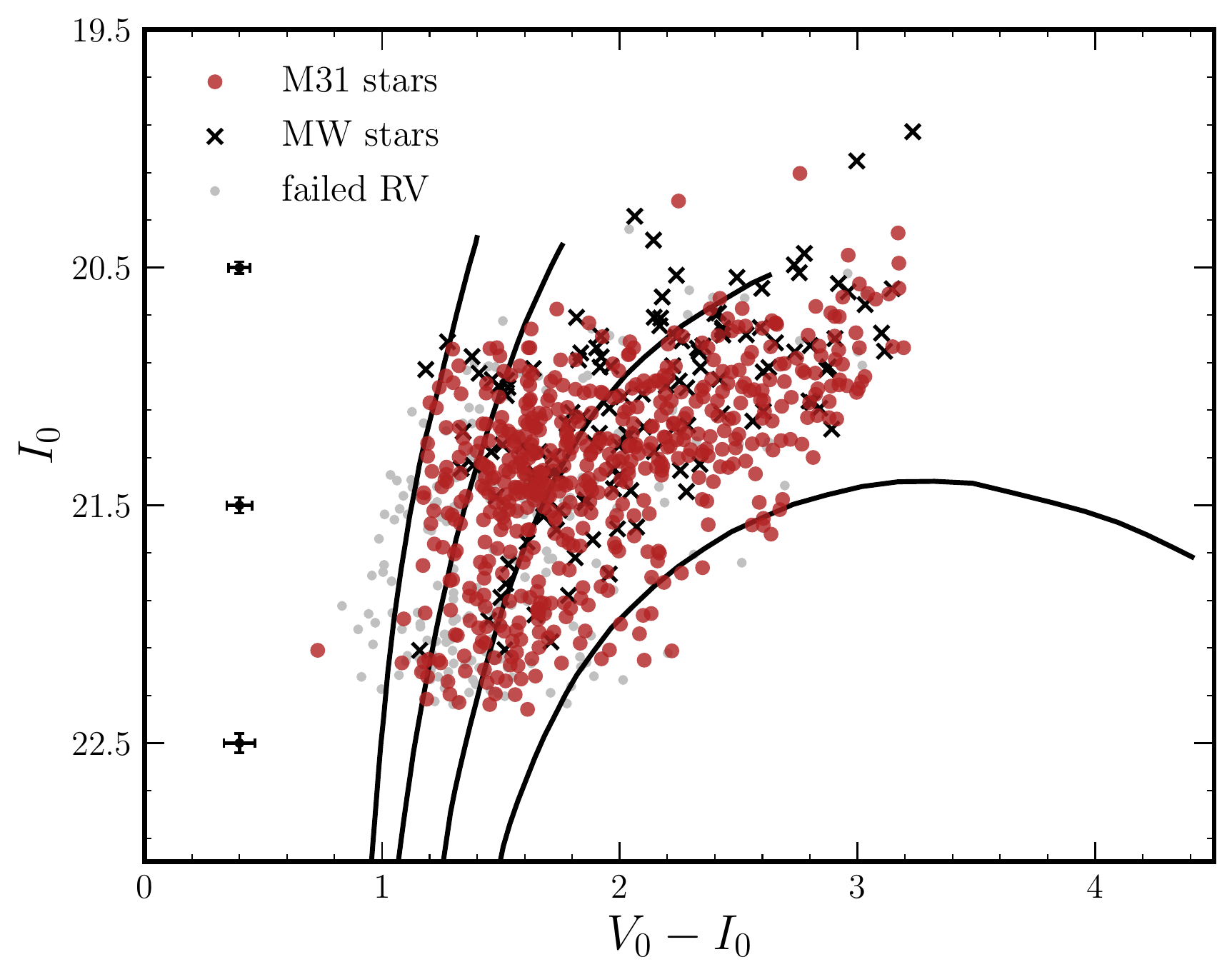}
    \caption{Extinction-corrected $(V_0,I_0)$ CMD (Section~\ref{sec:phot}) for all stars  in the NE shelf fields (Table~\ref{tab:obs}), including stars without successful radial velocity measurements (gray points), MW foreground stars (black crosses; Section~\ref{sec:mem}), and M31 RGB stars (red circles). The median photometric uncertainties as a function of magnitude are shown as black errorbars on the left.
    For reference, we show PARSEC isochrones (black lines; \citealt{Marigo2017}), assuming 12 Gyr ages and \alphafe\ = 0, with \feh\ = $-2.0$, $-1.0$, $-0.5$, and $0$ (from left to right).
    }
    \label{fig:cmd}
\end{figure}

\begin{figure*}
    \centering
    \includegraphics[width=\textwidth]{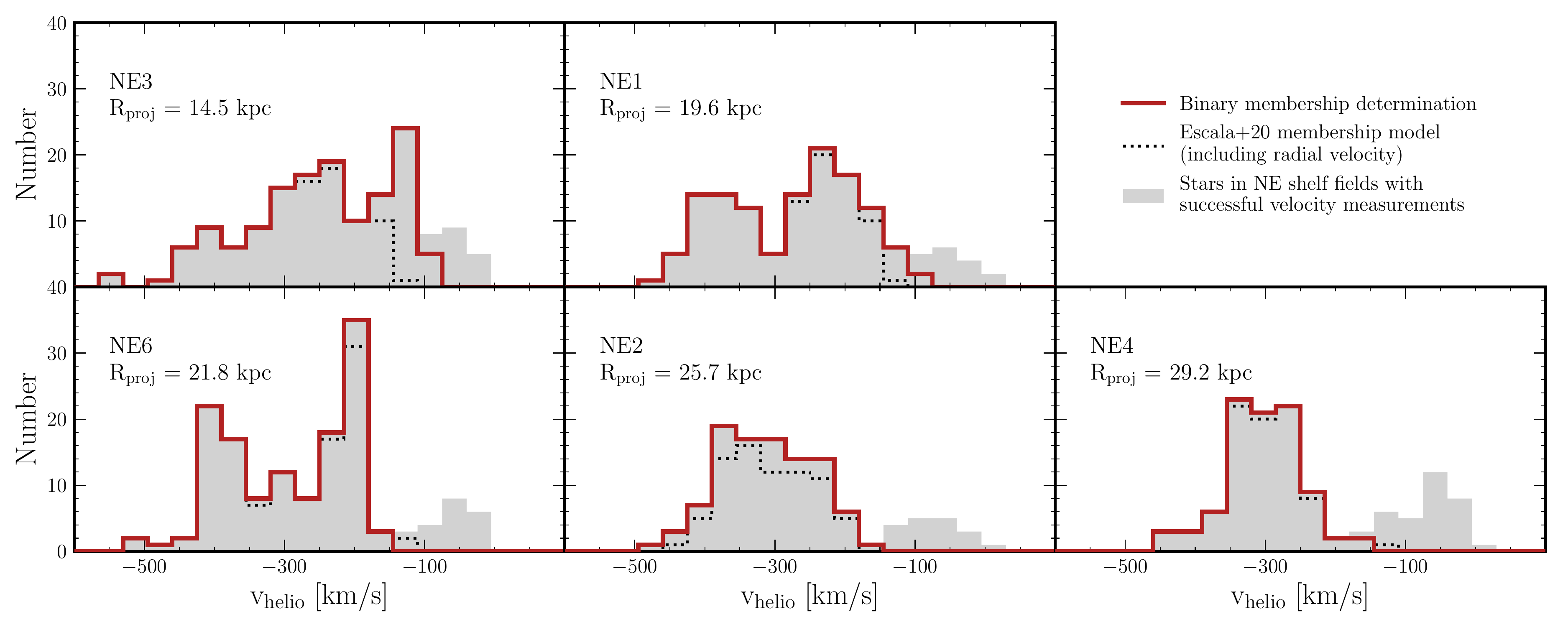}
    \caption{Heliocentric radial velocity distributions for stars with successful velocity measurements (grey filled histograms; Section~\ref{sec:obs}) in spectroscopic fields targeting the NE shelf (Table~\ref{tab:obs}), including both M31 RGB stars and MW foreground dwarf stars. From top to bottom and left to right, the fields probe larger M31-centric projected radii, approaching the ``tip'' of the tidal shell pattern (Section~\ref{sec:wedge}). The black dashed lines show the velocity distribution of stars classified as likely M31 members by the probabilistic model of \citet{Escala2020b}, which includes radial velocity as a membership diagnostic. This model mistakenly classifies NE shelf stars in field NE3 with MW-like velocities (\vhelio\ $\gtrsim -150$ \kms; Section~\ref{sec:mem}) as likely MW contaminants. Thus, we assigned stars with \vhelio\ $> -100$ ($-150$) \kms\ for the inner (outer) fields or spectral signatures of strong Na I $\lambda$8190 absorption as MW contaminants (red open histograms; Section~\ref{sec:mem}).}
   \label{fig:vhelio}
\end{figure*}

We sourced photometry from MegaCam images obtained in the $g^\prime, i^\prime$ bands using the 3.6 m Canada-France-Hawaii Telescope (CFHT). The images were reduced with the CFHT MegaPipe pipeline \citep{Gwyn2008} and the $g^\prime$, $i^\prime$ magnitudes were transformed to Johnson-Cousins $V,I$ using observations of Landolt photometric standard stars \citep{Kalirai2006}. When designing slitmasks, we identified RGB candidates based on only the $I$-band magnitude to avoid introducing bias into the target selection. This still yields low MW contamination (Section~\ref{sec:mem}) in the NE shelf region owing to the high density of M31 stars relative to the foreground below the tip of the RGB at $I_0\sim20.5$ magnitudes.

Figure~\ref{fig:cmd} shows the de-reddened $V_0,I_0$ CMD for all stars with successful radial velocity measurements (Section~\ref{sec:obs}) in the NE shelf fields (Table~\ref{tab:obs}). We corrected the photometry for the effects of dust extinction by assuming field-specific interstellar reddening values
from the maps of \citet{Schlegel1998} and applying the corrections defined by \citet{SchlaflyFinkbeiner2011}. We also show PARSEC isochrones \citep{Marigo2017} for reference, assuming an age of 12 Gyr, \alphafe\ = 0, and $m-M$ = \dm\ for consistency with previous studies in the SPLASH collaboration. Based on this grid of theoretical isochrones, we determined \fehphot\ via interpolation given the color and magnitude of each star \citep{Escala2020a}.

\section{Properties of the Northeast Shelf}
\label{sec:props}

\subsection{Membership}
\label{sec:mem}

In prior studies of M31's stellar halo, we employed likelihood-based methods \citep{Gilbert2006,Escala2020b} to separate M31 RGB stars from the MW foreground using diagnostic measurements such as radial velocity (\vhelio), Na I $\lambda$8190 doublet equivalent width (\ewna), photometric metallicity (\fehphot), and calcium-triplet based metallicity (\fehcat). In particular, the technique of \citet{Escala2020b} relies on Bayesian inference to assign each target star a probability of M31 membership based on the observed properties of over a thousand M31 and MW stars securely identified in SPLASH  \citep{Gilbert2006,Gilbert2012}. As a consequence, these methods were calibrated to the properties of the stellar halo in M31's southeast quadrant, where known M31 substructure is not present at MW-like heliocentric velocities (\vhelio\ $\gtrsim -150$ \kms; \citealt{Gilbert2018}). However, these approaches to membership determination overestimate the degree of MW contamination for the NE shelf fields when using radial velocity as a diagnostic. 

\begin{figure}
    \centering
    \includegraphics[width=\columnwidth]{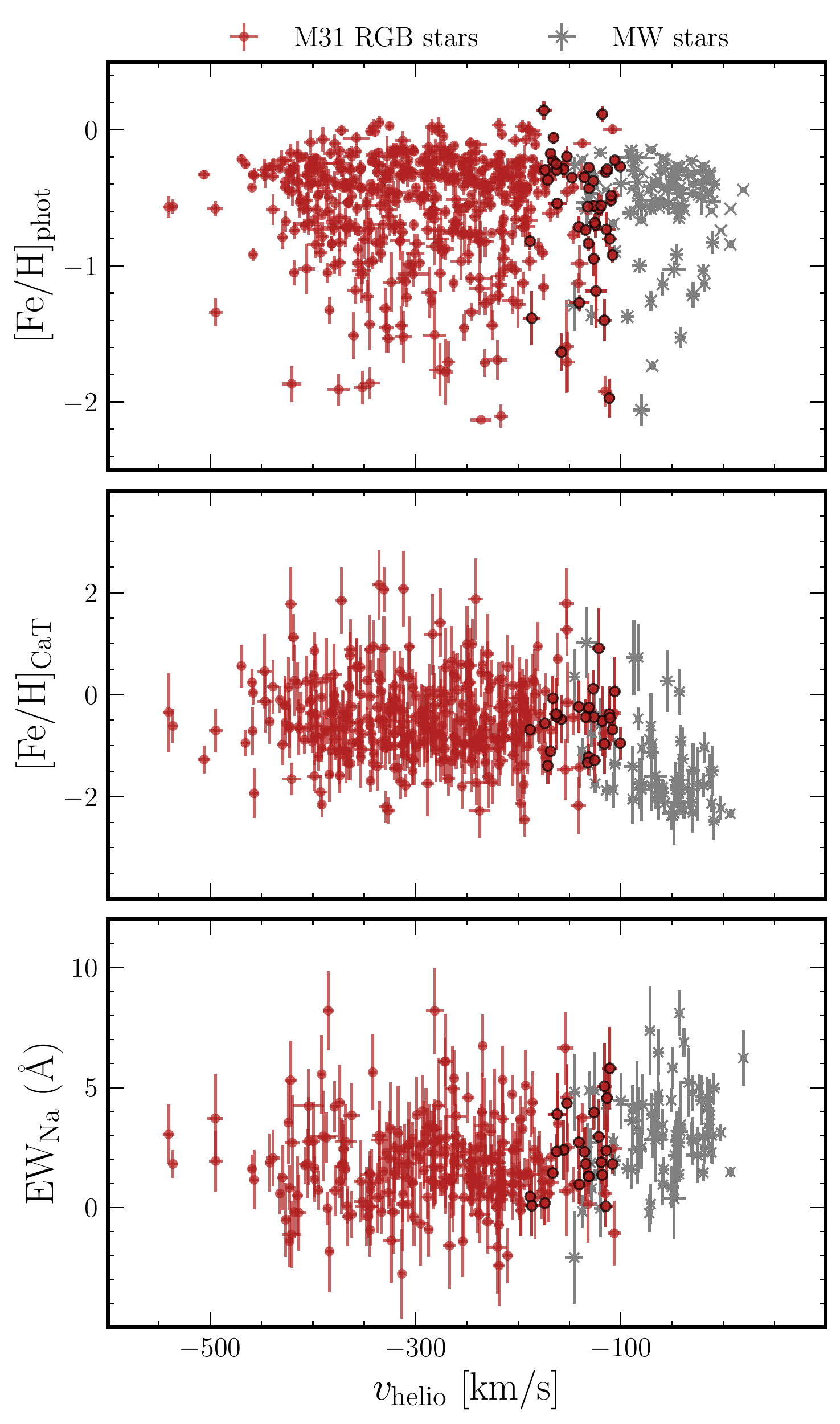}
    \caption{Photometric metallicity (\fehphot; top), spectroscopic calcium-triplet based metallicity (\fehcat; middle), and the summed equivalent width of the Na I $\lambda$8190 doublet (\ewna; bottom) versus heliocentric velocity for stars in the NE shelf fields (Table~\ref{tab:obs}) classified as M31 RGB stars (red circles) and MW stars (gray crosses; Section~\ref{sec:mem}). Stars in the NE3 velocity peak (Figure~\ref{fig:vhelio}) are outlined in black. We show \fehcat\ (\ewna) measurements
    with uncertainties \fehcaterr\ $<$ \fehcaterrmax\ dex (\ewnaerr\ $<$ \ewnaerrmax\ \AA), where the large uncertainties are due to the low spectral S/N. Neither \fehphot\ nor \fehcat\ are physically meaningful for MW stars but are useful as population diagnostics.
    }
    \label{fig:mem}
\end{figure}

For context, Figure~\ref{fig:vhelio} shows the heliocentric radial velocity distributions for stars in the NE shelf fields with successful velocity measurements, including both M31 RGB stars and intervening MW dwarf stars. 
In the innermost field {\tt NE3}, the \citeauthor{Escala2020b} model assigns a low probability of M31 membership to a prominent kinematical feature at \vhelio\ $\sim$ $-120$ \kms\ (Figure~\ref{fig:vhelio}) when including radial velocity as a diagnostic. However, this feature likely corresponds to the NE shelf based on its position and velocity (Section~\ref{sec:wedge}). When selecting stars in {\tt NE3} that have velocities consistent with this feature ($-190$ \kms\ $<$ \vhelio\ $< -100$ \kms), we found that their \ewna\ and \fehcat\ measurements
and CMD positions are consistent with those of stars in the NE shelf fields with M31-like velocities (\vhelio\ $\lesssim$ $-$200 \kms; Figure~\ref{fig:mem}). Thus, the likelihood-based M31 membership models are not uniformly applicable to the NE shelf fields when using radial velocity criteria due to the distinct kinematical structure of the NE shelf.



We adopted a binary membership determination in which we classified stars with clear spectral signatures of strong Na I $\lambda$8190 absorption or \vhelio\ $> -100$ \kms\ (\vhelio\ $> -150$ \kms) for \Rproj\ $<$ 20 kpc (\Rproj\ $>$ 20 kpc) as belonging to the MW foreground. We relied on visual inspection of the Na I doublet as opposed to \ewna\ measurements owing to the low S/N of the spectra. We do not detect stars with strong Na I absorption at M31-like velocities (Figure~\ref{fig:mem}) given that MW contaminants in this velocity range are likely blue main sequence turn-off stars in the MW's stellar halo.

The change in velocity cut with projected radius is a consequence of the shifting of the shelf velocity distribution relative to the constant MW foreground (Figure~\ref{fig:vhelio}). The velocity cut includes the majority of stars in the {\tt NE3} velocity peak 
while excluding the majority of stars at MW-like velocities in this field. 
We characterized the MW velocity distribution over the NE shelf region as probed by our selection function by utilizing an expectation-maximization algorithm to fit a 3-component Gaussian mixture (\citealt{sklearn}) to the combined velocity distribution of the three outermost fields. We found $\mu_{v,{\rm MW}} = $ \muvmw\ \kms, $\sigma_{v,{\rm MW}} = $ \sigvmw\ \kms, indicating that a velocity cut at $-$100 \kms\ removes \percentmwvel\ of stars at MW-like velocities when assuming a normal distribution.
In the outermost fields, the velocity cut should largely eliminate MW contamination, where \citet{Gilbert2007} found 
that MW dwarf stars along the line-of-sight to M31 are primarily limited to \vhelio\ $> -150$ \kms.

Figure~\ref{fig:mem} shows \fehphot, \fehcat,  and \ewna\ versus \vhelio\ for stars classified as belonging to M31 and the MW in the NE shelf fields.\footnote{Similar to \ewna, an equivalent width measurement for the calcium triplet in an individual star is subject to noise owing to the low S/N of the spectra. However, the statistical trends of \ewna\ and \fehcat\ are useful for distinguishing between the different populations of M31 and MW stars.} The \fehphot\ derivation (Section~\ref{sec:phot}) assumes that stars are at the distance of M31 and the \fehcat\ calibration is appropriate only for RGB stars, therefore both are inaccurate for MW stars. We emphasize that we did not use selections on \fehphot, \fehcat, or \ewna\ measurements to determine membership. Instead, Figure~\ref{fig:mem} illustrates that a binary membership classification based solely on \vhelio\ and apparent Na I absorption captures statistical differences between the observed properties of M31 and MW stellar populations, despite favoring completeness over lower contamination. For example, the median \fehcat\ and \ewna\ of M31 (MW) stars is \medfehcatand\ (\medfehcatmw) and \medewnaand\ (\medewnamw) \AA\ based on measurements with uncertainties $\delta$\fehcat\ $<$ \fehcaterrmax\ and $\delta$\ewna\ $<$ \ewnaerrmax\ \AA, respectively, which are consistent with known properties of M31 (MW) stars along the line-of-sight to M31 \citep{Gilbert2006,Escala2020b}. Table~\ref{tab:obs} provides the number of M31 RGB stars identified in each field. 

\subsection{Empirical Modeling of the Velocity Distributions}
\label{sec:vel_model}

\begin{figure*}
    \centering
    \includegraphics[width=\textwidth]{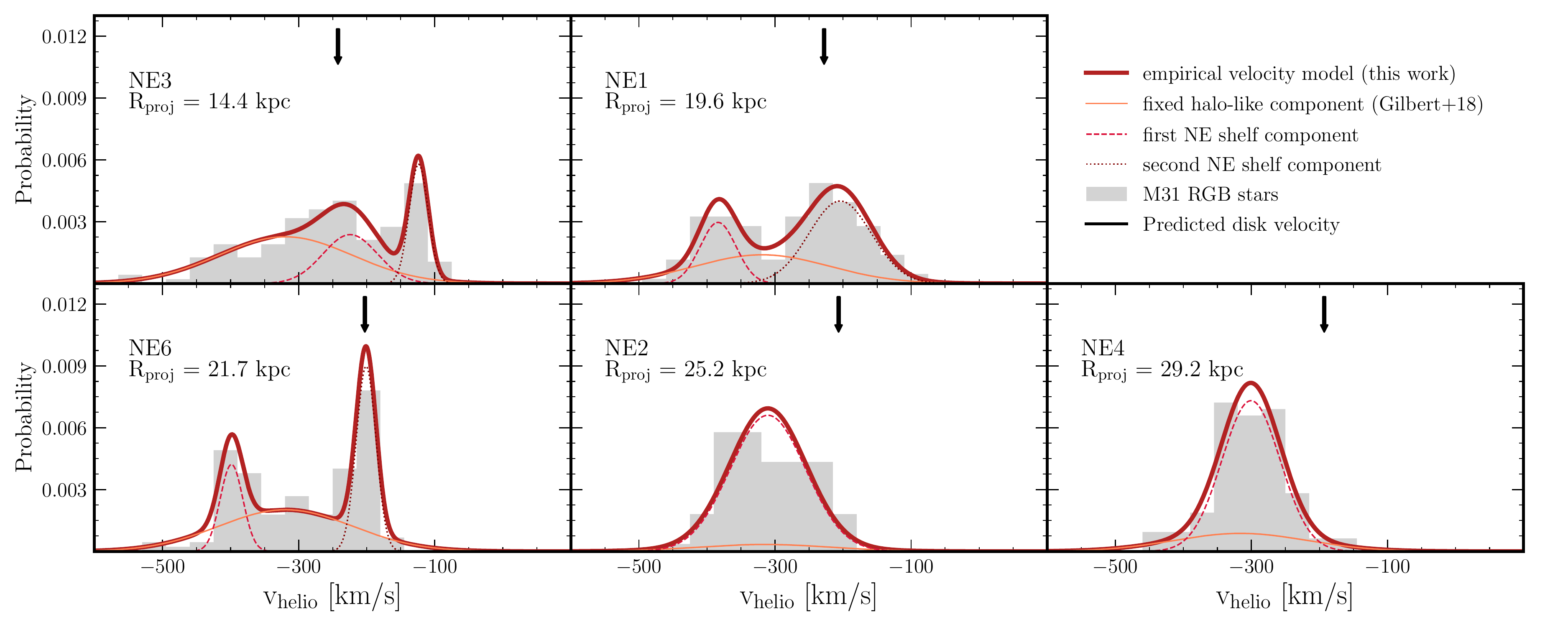}
    \caption{Heliocentric radial velocity distributions for M31 RGB stars (gray histograms; Section~\ref{sec:mem}) in the NE shelf fields. 
    We show the empirical velocity model derived for each field from the 50$^{\rm th}$ percentile values of the marginalized posterior probability distributions (thick red lines; Section~\ref{sec:vel_model}, Table~\ref{tab:vmodel}). We constructed the velocity models from a kinematically hot halo-like component with fixed mean and standard deviation (thin solid orange lines; \citealt{Gilbert2018}) and kinematically cold component(s) corresponding predominantly to NE shelf substructure (thin dotted and dashed red lines). We also show the median velocity of the disk predicted for each field assuming its presence (black arrows; Section~\ref{sec:disk}).
    }
    \label{fig:vmodel}
\end{figure*}

The observed velocity distributions (Figure~\ref{fig:vhelio}) clearly demonstrate that kinematical substructure is present across the spectroscopic fields. Such substructure is presumably associated with the tidal debris that constitutes the NE shelf. Thus, we performed {\it empirically} motivated modeling of the velocity distribution to identify probable substructure components directly from the data. This is in contrast to the N-body modeling used to predict the velocity structure of the NE shelf in Section~\ref{sec:nbody}.

\subsubsection{Number of Kinematical Components}
\label{sec:ncomp}

We confirmed that each field contains kinematical substructure by testing whether the velocity distributions for M31 RGB stars (Section~\ref{sec:mem}) are consistent with a pure stellar halo component. We assumed that the stellar halo in the NE shelf region can be described by the kinematically unbiased parameterization found by \citet{Gilbert2018} for M31's southeast quadrant via modeling the line-of-sight velocity distribution of over 5000 stars in radial bins spanning 9--175 projected kpc. 
We transformed the mean halo velocities in a given radial bin from the Galactocentric to heliocentric frame based on the median R.A. and Decl. of all stars in each NE shelf field. 
We performed an Anderson-Darling test with 10$^3$ Monte Carlo trials to compare the observed velocity distribution of each field to the accompanying Gaussian stellar halo model. In each iteration, we randomly drew samples from the model according to the number of M31 members in each field (Table~\ref{tab:obs}) and calculated the test statistic for a perturbation of the line-of-sight velocities assuming Gaussian measurement errors. We found that the velocity distribution for each field---including {\tt NE2} and {\tt NE4} despite the apparent lack of multiple velocity peaks---is highly inconsistent with a single kinematically hot stellar halo component. 

To determine the number of kinematical components in each field, we used an expectation-maximization (EM) algorithm to fit Gaussian mixtures \citep{sklearn} to the observed velocity distribution of M31 RGB stars. For each $N$-component model, we computed the Bayesian information criterion (BIC) for $10^3$ perturbations of the velocities according to their errors. 
We found that $N$-components = (2, 1, 3, 1, 2) minimizes the BIC for fields ({\tt NE1}, {\tt NE2}, {\tt NE3}, {\tt NE4}, {\tt NE6}). However, the EM algorithm has no knowledge of the structural properties of the substructure-dominated NE shelf region and likely disregards the dynamically hot stellar halo that must be present at some level in each field. We therefore incremented the number of components by one to account for a kinematically hot halo-like component in each field except {\tt NE3}. 

As demonstrated by Figure~\ref{fig:vmodel}, {\tt NE3} has a more complex velocity distribution, which makes the separation of substructure from a background halo-like component less straightforward. We evaluated whether {\tt NE3} contains 3 or 4 kinematical components when including a halo-like component (i.e., whether the field has 2 or 3 kinematically {\it cold} components) by modeling its velocity distribution using the procedure detailed in Section~\ref{sec:mcmc}. We calculated a distribution of BICs using the models described by every 100$^{\rm th}$ sampled parameter combination from the converged portion of the flattened Markov chain assuming 3 and 4 components. Based on this, we concluded that a 3 component model composed of 2 substructure components and a halo-like component provides a better representation of the data for this field.

\subsubsection{Sampling the Posterior Probability Distributions}
\label{sec:mcmc}

\begin{table*}
    \centering
    \caption{Empirical Velocity Distribution Model Parameters for NE Shelf Fields}
    \begin{threeparttable}
    \begin{tabular*}{\textwidth}{@{\extracolsep{\fill}}lccccccccc}
    \hline\hline
    Field & \multicolumn{1}{p{1.0cm}}{\centering R$_{\rm proj}$\\(kpc)} & \multicolumn{1}{p{1.0cm}}{\centering $\mu_{\rm halo}$\\(km/s)} & \multicolumn{1}{p{1.0cm}}{\centering $\sigma_{\rm halo}$\\(km/s)} & \multicolumn{1}{p{1.0cm}}{\centering $\mu_{\rm KCC1}$\\(km/s)} & \multicolumn{1}{p{1.0cm}}{\centering $\sigma_{\rm KCC1}$\\(km/s)} & \multicolumn{1}{p{1.0cm}}{\centering $f_{\rm KCC1}$\\(km/s)} & \multicolumn{1}{p{1.0cm}}{\centering $\mu_{\rm KCC2}$\\(km/s)} & \multicolumn{1}{p{1.0cm}}{\centering $\sigma_{\rm KCC2}$\\(km/s)} & \multicolumn{1}{p{1.0cm}}{\centering $f_{\rm KCC2}$\\(km/s)}\\ \hline
    NE3 & 14.4 & $-$319.4 & 98.1 & $-$224.3$^{+20.3}_{-18.0}$ & 40.0$^{+14.1}_{-15.4}$ & 0.24$^{+0.11}_{-0.1}$ & $-$123.7$^{+3.9}_{-6.1}$ & 14.0$^{+4.9}_{-2.9}$ & 0.20$^{+0.05}_{-0.04}$\\
    NE1 & 19.6 & $-$319.0 & 98.1 & $-$383.7$^{+11.2}_{-10.2}$ & 26.4$^{+9.7}_{-9.0}$ & 0.20$^{+0.11}_{-0.09}$ & $-$204.2$^{+11.9}_{-10.5}$ & 45.9$^{+7.7}_{-7.7}$ & 0.46$^{+0.11}_{-0.12}$\\
    NE6 & 21.7 & $-$319.1 & 98.1 & $-$398.6$^{+7.1}_{-6.3}$ & 16.6$^{+4.7}_{-3.8}$ & 0.18$^{+0.05}_{-0.05}$ & $-$200.9$^{+3.0}_{-3.1}$ & 14.5$^{+2.4}_{-2.1}$ & 0.33$^{+0.05}_{-0.05}$\\
    NE2 & 25.2 & $-$316.2 & 98.0 & $-$310.5$^{+7.3}_{-7.4}$ & 55.4$^{+5.2}_{-5.1}$ & 0.92$^{+0.06}_{-0.14}$ & ... & ... & ...\\
    NE4 & 29.2 & $-$316.0 & 98.0 & $-$300.6$^{+6.2}_{-6.3}$ & 42.8$^{+7.3}_{-6.3}$ & 0.79$^{+0.12}_{-0.13}$ & ... & ... & ...\\
    \hline
    \end{tabular*}
    \begin{tablenotes}[flushleft]
    \footnotesize
    \item Note. \textemdash\ The parameters describing the model components are mean velocity ($\mu$), velocity dispersion ($\sigma$), and normalized fractional contribution ($f$). We constructed each model from halo-like \citep{Gilbert2018} and kinematically cold component(s) (KCCs) corresponding predominantly to NE shelf substructure (Section~\ref{sec:vel_model}). The parameter values are the 50$^{\rm{th}}$ percentiles of the marginalized posterior probability distributions, where the errors are calculated from the 16$^{\rm{th}}$ and 84$^{\rm{th}}$ percentiles.
    \end{tablenotes}
    \end{threeparttable}
    \label{tab:vmodel}
\end{table*}


We modeled the observed velocity distribution in each field as a Gaussian mixture composed of kinematically cold components(s) that correspond to the NE shelf (see Section~\ref{sec:disk} for a discussion on the disk) and a kinematically hot halo-like component with fixed mean and dispersion but variable fractional contribution. We refer to the kinematically hot component as ``halo-like'' to emphasize that it may be composed of both a classical stellar halo (i.e., in-situ and phase-mixed accreted stars) and high-velocity-dispersion tidal debris constituting part of the shell. The log likelihood of this model is given by,

\begin{equation}
\label{eq:lkhd}
\ln \mathcal{L} = \sum_{i=1}^{N} \ln \left( \sum_{k=1}^{K} f_k \mathcal{N} (v_i | \mu_k, \tau_k^{-1}) \right),
\end{equation}
where $i$ is the index representing a RGB star, $v_i$ is its heliocentric radial velocity, and $N$ is the number of RGB stars. $K$ is the total number of components (including the kinematically hot component), $k$ is the index for a given component, and $\mu_k$, $\tau_k = \sigma_k^{-2}$, and $f_k$ are the mean, inverse variance, and fractional contribution of each component. Note that the fractional contribution of the halo-like component is not a free parameter and is instead constrained by the kinematically cold component(s) (KCCs).

We  sampled  from  the  posterior distribution  of  the velocity  model  (Eq.~\ref{eq:lkhd})  for  each  field  using  an  affine-invariant  Markov chain  Monte  Carlo  (MCMC)  ensemble sampler ({\tt emcee}; \citealt{Foreman-Mackey2013}) with 10$^2$ walkers and $10^4$ steps. We limited the range of our model parameters to samples drawn from $-$600 \kms\ $< \mu_k <$ $-$100 \kms, 5 \kms\ $<$ $\sigma_k$ $<$ 150 \kms, and $0 < f_k < 1$, and required that $\mu_k < \mu_{k+1}$ and $\sum f_k \leq 1$ for KCCs. We implemented uniform priors on $\mu_k$ over the allowed range for maximal flexibility given the previously unstudied nature of the NE shelf velocity distribution for RGB stars. We assumed Gamma priors on $\tau_k$ with $\alpha$ = \alphagamma\ and $\beta$ = \betagamma\ to penalize small values of $\sigma_k$ while enabling a lower probability extended tail toward high values of $\sigma_k$. We constructed the marginalized posterior distributions from the latter 50\% of the samples, where Table~\ref{tab:vmodel} summarizes the model parameters calculated from the 16$^{\rm th}$, 50$^{\rm th}$, and 84$^{\rm th}$ percentiles of the posterior distributions.

Figure~\ref{fig:vmodel} shows the heliocentric velocity distributions for RGB stars in each field compared to its adopted velocity model. We also indicate the separate kinematically hot and cold components constituting the model in the figure.  As previously discussed, the interior field {\tt NE3} has the most complicated velocity distribution, which is composed of a cold feature at \vhelio\ $\sim$ $-120$ \kms\ that is likely the upper envelope of the NE shelf (Section~\ref{sec:wedge} and~\ref{sec:nbody}), a significant halo-like component, and a third component of unclear origin (c.f. Section~\ref{sec:disk}). The fields at intermediate projected radii, {\tt NE1} and {\tt NE6}, consist of two NE shelf components corresponding to the envelopes of the tidal shell (Section~\ref{sec:wedge} and~\ref{sec:nbody}). The outermost fields, {\tt NE2} and {\tt NE4}, are dominated by NE shelf substructure, where relatively insignificant halo-like components are favored by the model. 

Based on the model for each field, we assigned each star a probability of belonging to substructure ($p_{\rm sub}$) given its heliocentric velocity. This probability is computed from the likelihood ratio between the KCC model(s) and halo-like component model. We emphasize that stars with lower values of $p_{\rm sub}$ may in fact belong to NE shelf substructure; $p_{\rm sub}$ should be interpreted as a measure of confidence that a star is associated with tidal debris based on velocity information alone.

\subsubsection{Potential Contamination from M31's Disk}
\label{sec:disk}

We assessed whether a significant contribution from M31's disk is expected in the NE shelf velocity distribution owing to its complex kinematical structure (Figure~\ref{fig:vmodel}) and the proximity of the fields to the disk (Figure~\ref{fig:fields}). Using a simple model for the perfectly circular rotation of an inclined disk \citep{Guhathakurta1988} with inclination angle $i=77^\circ$ and position angle P.A. = 38$^\circ$, we calculated the predicted line-of-sight velocities for a disk feature over the R.A., Decl. range (corresponding to \Rdisk\ = 28--42 kpc) spanned by our fields. Based on a rotation curve derived from HI kinematics out to \Rdisk\ = 38 kpc and corrected for the inclination of the disk \citep{Chemin2009}, the expected disk rotation velocity in the NE shelf fields is 
$\sim$250 \kms. 

When using the simple model, this translates to line-of-sight velocities of $v_{\rm disk, med}$ = \vdisknethree, \vdiskneone, \vdisknesix, \vdisknetwo, and \vdisknefour\ \kms\ in fields {\tt NE3}, {\tt NE1}, {\tt NE6}, {\tt NE2}, and {\tt NE4}, respectively.
We note that the line-of-sight velocity dispersion of the disk is $\sim$50-60 \kms\ as found by \citet{Dorman2012} using a similar velocity modeling methodology (Section~\ref{sec:vel_model}) over a comparable radial range to the NE shelf fields. Moreover, the rotation velocity of RGB stars lags the HI disk by \rgbad\ \kms\ \citep{Quirk2019}, such that adjusting for this asymmetric drift would decrease $v_{\rm disk, med}$ by $\sim$15--30 \kms.

In contrast to fields {\tt NE2} and {\tt NE4}, fields {\tt NE1}, {\tt NE3}, and {\tt NE6} show evidence of kinematically cold components near $v_{\rm disk, med}$ (Figure~\ref{fig:vmodel}; Table~\ref{tab:vmodel}). However, the majority of stars in the KCC at disk-like velocities in {\tt NE6} likely belong to the NE shelf given that its position in projected phase space corresponds to upper envelope of the shell pattern
(Section~\ref{sec:wedge} and~\ref{sec:nbody}). The KCC in {\tt NE3} at \vhelio\ $\sim$ $-$220 \kms\ could feasibly originate from the disk, whereas the KCC at \vhelio\ $\sim$ $-200$ \kms\ in {\tt NE1} may be contaminated by the disk. We further explore these features in comparison to predictions of N-body models in Section~\ref{sec:nbody}.

\subsection{Photometric Metallicity Distribution}
\label{sec:phot_metal}

\begin{figure}
    \centering
    \includegraphics[width=\columnwidth]{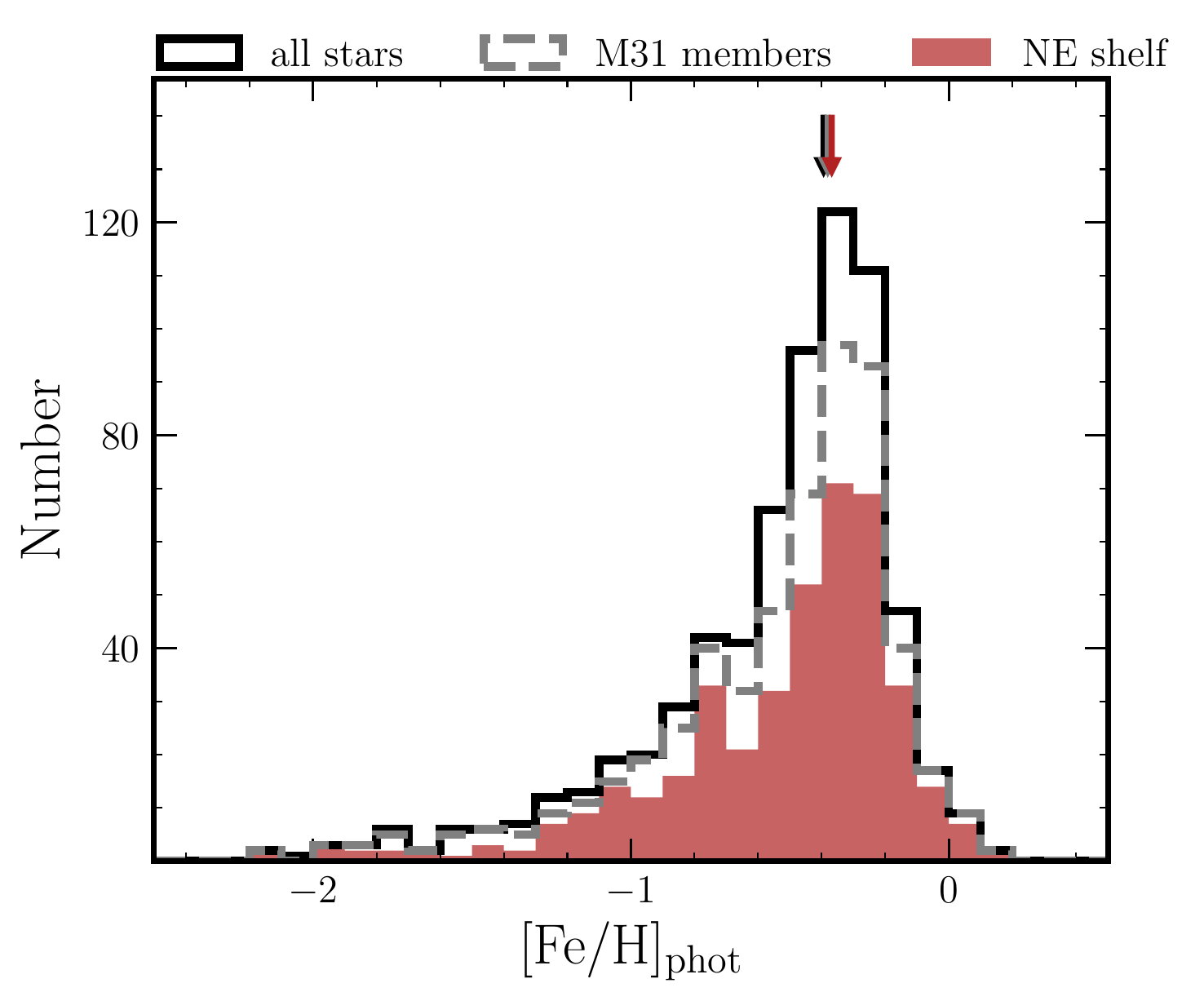}
    \caption{Photometric metallicity (\fehphot) distribution functions (MDFs) for all stars in the NE shelf fields (including MW contaminants; black solid open histogram), stars classified as M31 members (grey dashed open histogram; Section~\ref{sec:mem}), and stars that probably belong to the NE shelf ($p_{\rm sub} > 0.5$; Section~\ref{sec:mcmc}). We assumed 12 Gyr PARSEC isochrones \citep{Marigo2017} to determine \fehphot\ (Section~\ref{sec:phot}). The median \fehphot\ for each sample is shown as arrows, where \fehphot$_{\rm, med}$ = \fehphotmedne\ is robust with respect to sample selection, indicating that the NE shelf is the dominant feature in the data (Section~\ref{sec:phot_metal}).
    }
    \label{fig:mdf}
\end{figure}

\begin{table}
 \centering
 \caption{Photometric Metallicity Distribution Properties for Kinematically Cold Components in the NE Shelf Fields}
\begin{threeparttable}
    \begin{tabular*}{\columnwidth}{lcccc}
    \hline \hline
    Comp. & \multicolumn{1}{p{1.0cm}}{\centering $\mu_{\rm KCC}$\\(km/s)} & [Fe/H]$_{\rm med}$ & $\langle$[Fe/H]$\rangle$  & $\sigma_{\rm [Fe/H]}$ \\
    \hline
    Halo & ... & $-0.43^{+0.01}_{-0.02}$ & $-0.56$ $\pm$ 0.02 & 0.39 $\pm$ 0.02 \\
    \hline
    \multicolumn{5}{c}{NE3}\\
    \hline
     KCC1 & $-224.3$ & $-0.51^{+0.06}_{-0.04}$ & $-0.61$ $\pm$  0.04 &  0.42 $\pm$ 0.03\\
     KCC2 & $-123.7$ & $-0.57^{+0.01}_{-0.02}$ & $-0.65$ $\pm$  0.04 &  0.42$^{+0.04}_{-0.03}$\\
    \hline
    \multicolumn{5}{c}{NE1}\\
    \hline
    KCC1 & $-383.7$ & $-0.39^{+0.00}_{-0.08}$ & $-0.53^{+0.03}_{-0.04}$ & 0.37 $\pm$ 0.04 \\
    KCC2 & $-204.2$ & $-0.40^{+0.03}_{-0.02}$ & $-0.49$ $\pm$  0.03 &  0.37 $\pm$ 0.04 \\
    \hline
    \multicolumn{5}{c}{NE6}\\
    \hline
    KCC1 & $-398.6$ & $-0.41^{+0.0}_{-0.02}$ & $-0.50$ $\pm$ 0.02  & 0.27 $\pm$ 0.02\\
    KCC2 & $-200.9$ & $-0.40^{+0.03}_{-0.02}$ & $-0.49$ $\pm$  0.03 &  0.34 $\pm$ 0.02 \\
    \hline
    \multicolumn{5}{c}{NE2}\\
    \hline
     KCC1 & $-310.5$ & $-$0.45$^{+0.03}_{-0.04}$ & $-0.52$ $\pm$ 0.04 & 0.35 $\pm$ 0.05 \\
    \hline
    \multicolumn{5}{c}{NE4}\\
    \hline
     KCC1 & $-300.6$ & $-0.35^{+0.02}_{-0.05}$ & $-0.52$ $\pm$ 0.05 & 0.44 $\pm$ 0.05 \\
    \hline \hline
    \end{tabular*}
    \begin{tablenotes}[flushleft]
    \footnotesize
    \item Note.\@ \textemdash\ The columns are kinematical component, mean velocity (Table~\ref{tab:vmodel}), median \fehphot, mean \fehphot, and standard deviation on \fehphot. We calculated each quantity via bootstrap resampling weighted by 
    the probability that a star belongs to a given kinematical component (Section~\ref{sec:mcmc}). The quantities for the halo-like population across all fields are calculated analogously using $p_{\rm halo} = 1 - p_{\rm sub}$.
    \end{tablenotes}
    \label{tab:phot}
\end{threeparttable}
\end{table}

\begin{figure*}
    \centering
    \includegraphics[width=\textwidth]{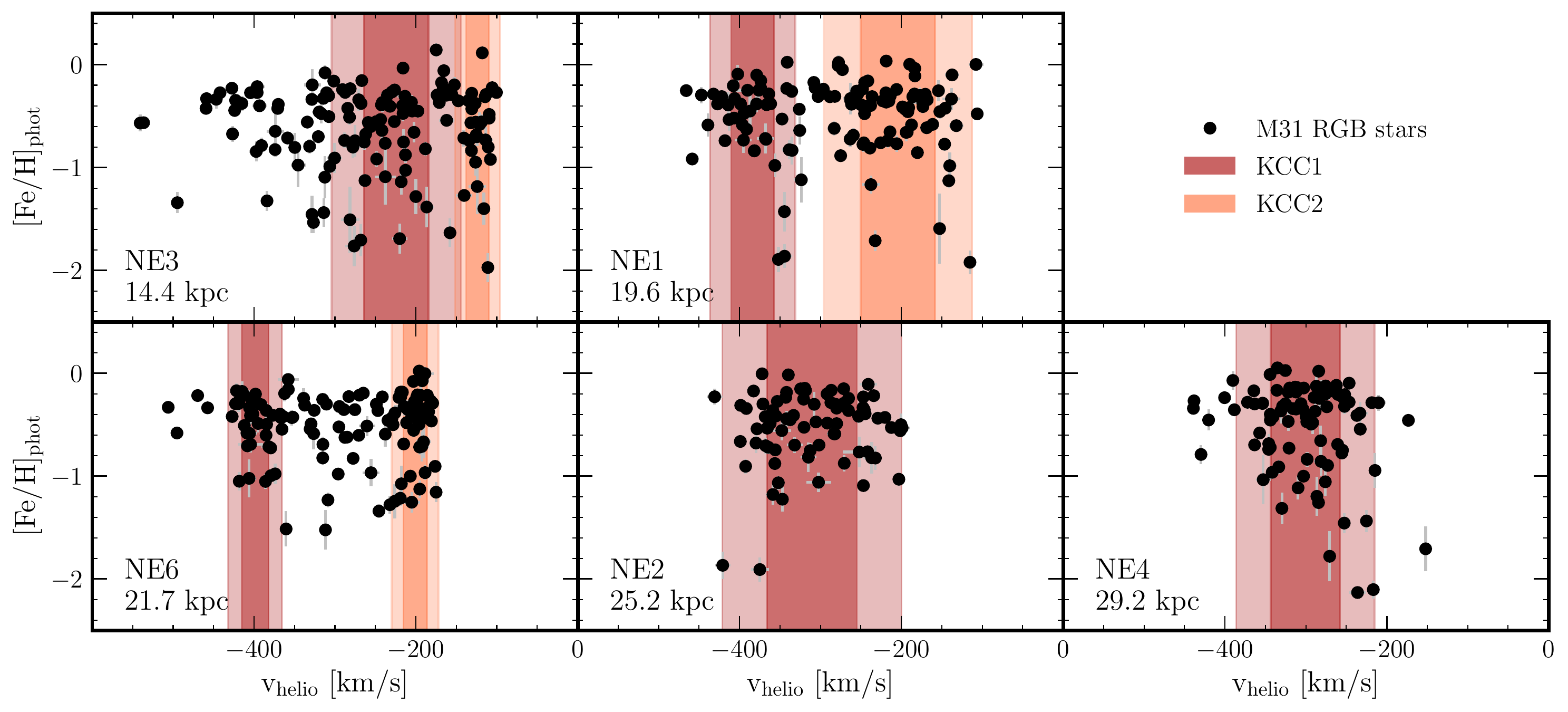}
    \caption{Heliocentric velocity (\vhelio) versus photometric metallicity (\fehphot) for M31 RGB stars (black points) in the NE shelf fields, where projected M31-centric radius (\Rproj) increases from left to right and top to bottom. We show the 1$\sigma$ (2$\sigma$) radial velocity range of KCC1 (red) and KCC2 (orange) (Table~\ref{tab:vmodel}; Figure~\ref{fig:vmodel}) as dark (light) shaded regions. The \fehphot\ distributions across the various NE shelf substructure components are similar. Stars assigned to the stellar halo component also have similar \fehphot\ to the NE shelf. 
    }
    \label{fig:vhelio_vs_fehphot}
\end{figure*}

We investigated the photometric metallicity distribution functions (MDFs) of RGB stars along the line-of-sight to the NE shelf to probe its chemical composition. We used photometric metallicities instead of spectroscopic calcium-triplet based metallicities (e.g., Figure~\ref{fig:mem}) because they are more robust at the low S/N characteristic of spectra in M31: \fehcat\ measurements are subject to larger random uncertainties than \fehphot\ measurements and are unavailable for a non-negligible number of stars. Moreover, we are concerned with relative metallicity differences between NE shelf stars and other stellar populations, as opposed to absolute metallicities. We determined \fehphot\ assuming 12 Gyr PARSEC isochrones \citep{Marigo2017} following Section~\ref{sec:phot}.

In Figure~\ref{fig:mdf}, we show the photometric MDFs for all stars in the NE shelf fields (including MW contaminants), stars classified as M31 members (Section~\ref{sec:mem}), and stars that probably belong to the NE shelf ($p_{\rm sub} > 0.5$; Section~\ref{sec:mcmc}). The unweighted median (mean) \fehphot\ for each sample is \fehphotmedall\ (\fehphotavgall), \fehphotmedmem\ (\fehphotavgmem), and \fehphotmedne\ (\fehphotavgne), respectively. If we instead assume 10 Gyr or 8 Gyr isochrones, which approximately correspond to the mean stellar age of M31's phase-mixed stellar halo \citep{Brown2006,Brown2007,Brown2008} and the GSS \citep{Brown2006,Tanaka2010}, we obtain a median (mean) \fehphot\ for the NE shelf that is \fehphotmeddiffhaloage\ (\fehphotavgdiffhaloage) dex or \fehphotmeddiffgssage\ (\fehphotavgdiffgssage) dex more metal-rich. The relative invariance of the median (mean) \fehphot\ value with respect to the sample selection indicates that NE shelf tidal debris is the dominant feature in the data. That is, the majority of stars with successful velocity measurements are indeed M31 RGB stars (Section~\ref{sec:mem}), although the value of \fehphot\ is not physically meaningful for MW foreground stars, and the majority of RGB stars indeed belong to the NE shelf and not the phase-mixed stellar halo.

In addition, the general shape of the MDF remains relatively constant between each sample, where the MDF is skewed toward the peak at \fehphot\ $\sim -0.4$, followed by a sharp decline for \fehphot\ $\gtrsim -0.2$ and an extended low-metallicity tail. The fact that the metal-poor (\fehphot\ $\lesssim -1.3$) population declines when selecting only for NE shelf stars ($p_{\rm sub} > 0.5$; Figure~\ref{fig:mdf}) suggests that the majority of these stars belong to M31's stellar halo rather than the shelf. Adopting stricter threshold for NE shelf membership ($p_{\rm sub} > 0.75$) does not significantly alter the location or shape of the MDF, so we retain the more inclusive criterion to maximize the sample size of NE shelf stars.

The most pronounced change in the MDFs in Figure~\ref{fig:mdf} is that a distinct peak at \fehphot\ $\sim$ $-0.7$ emerges in the NE shelf sample. This feature is robust because the bin sizes (0.1 dex) for each MDF are twice the median photometric uncertainty (\fehphoterrmed\ dex). This peak may originate from our choice to designate stars that probably belong to kinematical substructure ($p_{\rm sub} > 0.5$) as NE shelf stars, when in reality some of these stars could belong to other dynamically cold structures such as M31's disk (Section~\ref{sec:disk},~\ref{sec:nbody_wedge}). Interestingly, this peak metallicity roughly corresponds to the median \fehphot\ of M31's disk edge in the Pan-chromatic Hubble Andromeda Treasury (PHAT; \citealt{Dalcanton2012,Williams2014}) when assuming 13 Gyr isochrones \citep{Gregersen2015}.

\begin{figure*}
    \centering
    \includegraphics[width=\textwidth]{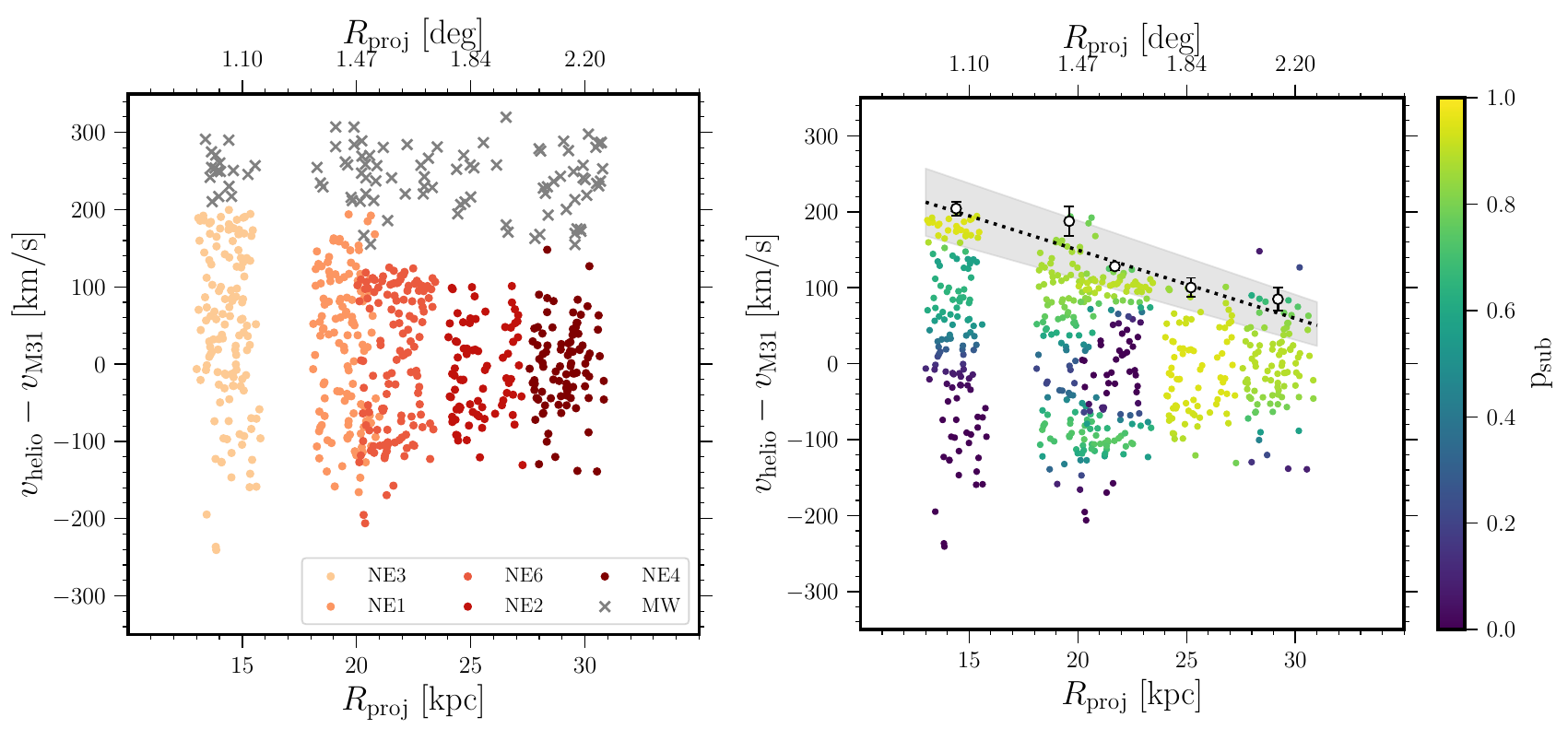}
    \caption{Heliocentric velocity relative to M31's systemic velocity ($v_{\rm M31}$ = \vsys\ \kms) versus projected M31-centric radius in the NE shelf fields for (left) stars with successful radial velocity measurements, including both M31 RGB stars (colored points) and MW foreground stars (grey crosses; Section~\ref{sec:mem}), and (right) solely M31 RGB stars color-coded by their kinematically-based probability of belonging to substructure (Section~\ref{sec:mcmc}). Velocity errors associated with individual data points are not shown for the sake of clarity (median \vtotmed\ \kms). The ``wedge'' pattern visible in this space for M31 RGB stars is characteristic of a tidal shell formed in a radial merger (Section~\ref{sec:wedge_obs}). The right panel also shows the observed upper envelope of the wedge (white open circles; Eq.~\ref{eq:up_env}) and its best-fit model (black dotted line, with 1$\sigma$ confidence interval as gray shaded region; \citealt{MerrifieldKuijken1998}) used to place constraints on M31's gravitational potential (Section~\ref{sec:wedge_model}).
    }
    \label{fig:wedge}
\end{figure*}

We further explored the \fehphot\ distribution of the NE shelf by considering the relationship to radial velocity and projected M31-centric radius. Figure~\ref{fig:vhelio_vs_fehphot} shows \vhelio\ versus \fehphot\ for each NE shelf field ordered as a function of \Rproj, where we highlighted the 2$\sigma$ radial velocity range associated with each detected kinematical substructure component (Table~\ref{tab:vmodel}; Figure~\ref{fig:vmodel}). 
We calculated the average metallicity and metallicity dispersion for each KCC weighted by the 
the kinematically-based probability of belonging to the given KCC (Table~\ref{tab:phot}). We note that the metallicity difference between kinematical components is likely larger than indicated by Table~\ref{tab:phot} because of the probabilistic nature of the component assignment. 
We found that the median \fehphot\ in {\tt NE3} is $\sim$0.15 dex more metal-poor than the relatively uniform metallicity of the other fields, possibly due to its more dominant halo-like population and/or larger disk contribution.
The halo-like population across all fields is similar in metallicity to the combined NE shelf population. This raises the possibility that this kinematically hot component may correspond to tidal debris with a high velocity dispersion, or similarly that the stellar halo in this region is heavily polluted by NE shelf substructure. We compare the metallicity distribution of the NE shelf to M31's phase-mixed stellar halo in Section~\ref{sec:comp_phot}.

Moreover, Figure~\ref{fig:vhelio_vs_fehphot} demonstrates that the aforementioned stars with \fehphot\ $\sim$ $-0.7$ dex are visible in the secondary KCC of {\tt NE1} and possibly the primary KCC of {\tt NE3}. These stars therefore have metallicities and velocities (Section~\ref{sec:disk}) broadly consistent with M31's disk, implying that a low level of disk contamination may be present in the data. However, the \fehphot\ distribution 
for stars that probably belong to KCCs (i.e., either the NE shelf or the disk) still appears to be dominated by the NE shelf, where removing stars with $-0.65 <$ \fehphot\ $< -0.85$ 
increases the median (mean) \fehphot\ by merely \fehphotmeddiffnenodisk\ (\fehphotavgdiffnenodisk) dex. The clump of stars at \fehphot\ $\lesssim$ $-1.0$ in the secondary KCC of {\tt NE6} could also be associated with the disk, where removing this grouping increases the median (mean) \fehphot\ by an additional \fehphotmeddiffnenodisksix\ (\fehphotavgdiffnenodisksix) dex.

\subsection{Projected Phase Space Distribution}
\label{sec:wedge}

Prior studies (e.g., \citealt{Fardal2007,Fardal2013,Sadoun2014,Miki2016,Kirihara2017}) have proposed that the NE shelf may have formed from a radial merger -- a merger with low angular momentum that consequently produces shell-like tidal structures \citep{HernquistQuinn1988,MerrifieldKuijken1998,SandersonHelmi2013}. These tidal structures create a ``caustic surface'' when viewed in radial ($r, v_r$) phase space, where this surface contains material with similar orbital energy that originates from the progenitor disrupted by the radial merger. Owing to the near-symmetry of this type of event, the resulting signature in projected ($R, v_{\rm los}$) phase space near the spatial edge of the shell depends only on the enclosed mass of the host galaxy and the shell radius \citep{MerrifieldKuijken1998,Ebrova2012,SandersonHelmi2013}. In this section, we provide an overview of the observed features of the projected phase space distribution for the NE shelf (Section~\ref{sec:wedge_obs}) before using the observed distribution to assess the ability of analytical models to constrain M31's gravitational potential at the shelf location (Section~\ref{sec:wedge_model}).

\subsubsection{Observed Features of the NE Shelf}
\label{sec:wedge_obs}

\begin{figure}
    \centering
    \includegraphics[width=\columnwidth]{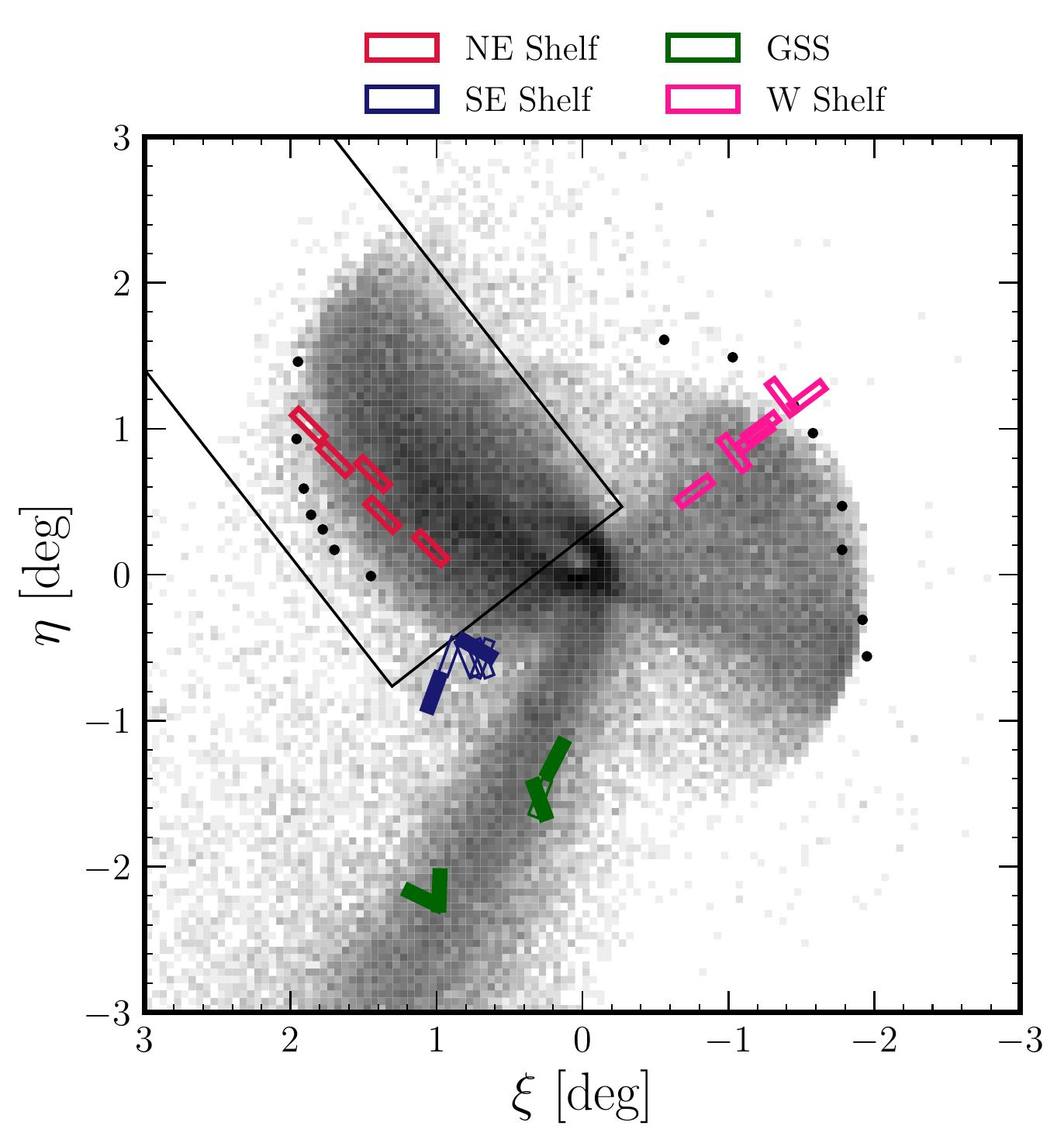}
    \caption{Histogram of the spatial distribution of satellite debris in from the F07 GSS formation model in M31-centric coordinates (Section~\ref{sec:nbody}). We show the spectroscopic field locations (colored rectangles) and photometric NE and W shelf edges (black points) from Figure~\ref{fig:fields}. The black outlined box is the broad selection region used to initially define the simulated NE shelf.
    }
    \label{fig:nbody_roadmap}
\end{figure}

\begin{figure*}
    \centering
    \includegraphics[width=\textwidth]{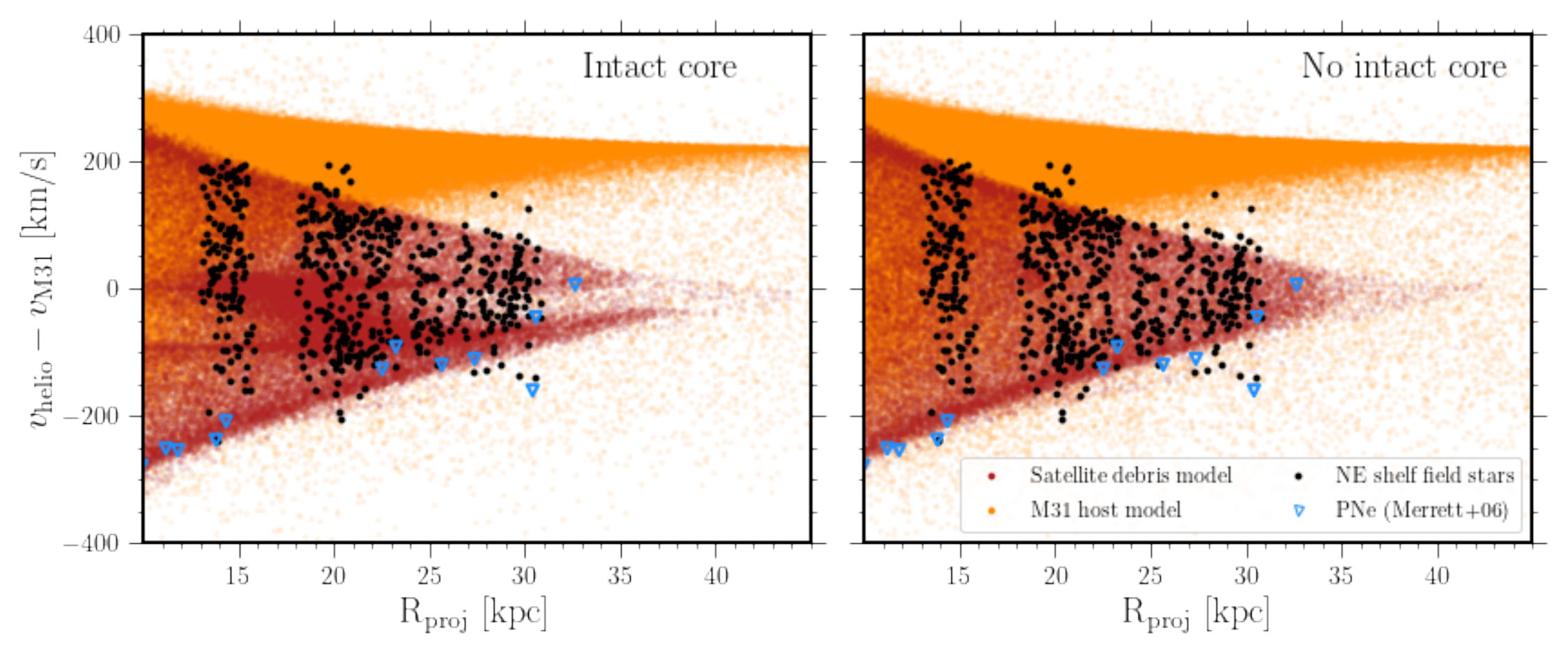}
    \caption{Similar to Figure~\ref{fig:wedge}, except comparing the predictions of N-body models \citep{Fardal2007,Fardal2012} in the NE shelf region for the formation of the GSS to observations of RGB stars (black points; this work) and PNe (blue open triangles; \citealt{Merrett2003,Merrett2006}). The models contain a combined bulge, disk, and halo component for the host (orange points) and a satellite component corresponding to GSS-related tidal debris (red points). We show models for a merger remnant with (left panel) and without (right panel) an intact core. The NE shelf region is defined using $X_{\rm M31} >$ 0.2$^\circ$ and $-1.5^\circ < Y_{\rm M31} < 0.5^\circ$, where $X_{\rm M31}$ and $Y_{\rm M31}$ are defined along the NE major axis (P.A.\@ = 38$^\circ$ E of N) and NW minor axis, respectively. 
    The observed projected phase space distribution of the NE shelf broadly agrees with that of a tidal shell formed from a radial merger with the GSS progenitor, where the model with a completely disrupted progenitor provides a better match (Section~\ref{sec:nbody_core}).
    }
    \label{fig:nbody_wedge}
\end{figure*}

Figure~\ref{fig:wedge} shows the line-of-sight velocity (\vhelio) of stars in the NE shelf fields, shifted according to M31's systemic velocity ($v_{\rm M31} = -300$ \kms), versus projected M31-centric radius (\Rproj). In this space, M31 RGB stars clearly show a ``wedge'' pattern characteristic of a shell formed in a radial merger: the \vhelio\ range decreases with increasing \Rproj\ as stars in the shell approach apocenter, where \vhelio\ $-$ $v_{\rm M31} \sim 0$ \kms. Given that the velocity dispersion of the NE shelf in the outermost field ({\tt NE4}; \Rproj\ $\sim$ 29 kpc) is relatively large ($\sigma_v \sim 43$ \kms\ ; Table~\ref{tab:vmodel}), 
the data may not capture the full radial extent of the shell feature (see Figure~\ref{fig:nbody_wedge}).\footnote{The simulations do not predict a distinct ``tip'' of the wedge pattern in projected phase space for the NE shelf (Figure~\ref{fig:nbody_wedge}), which suggests that the NE shelf may not have a well-defined spatial edge regardless of the spatial coverage of the spectroscopic data.}

The magnitude difference between red clump stars in the NE shelf and GSS indicates
that it is located in front of the latter feature given their similar CMD morphologies \citep{Ferguson2005,Richardson2008} and consequently that the NE shelf is located in front of M31 when combined with line-of-sight distance measurements \citep{McConnachie2003}. Thus, the upper envelope (\vhelio $-$ $v_{\rm M31} > 0$ \kms) of the caustic corresponds to tidally stripped stars moving toward M31, whereas the lower envelope corresponds to stars moving away from M31. 
In the field with the innermost robust detection of both envelopes of the wedge ({\tt NE1}; Figure~\ref{fig:vmodel}), the positive (negative) envelope has \vhelio\ $-$ $v_{\rm M31}$ $\sim +95.8^{+11.9}_{-10.5}$ ($-83.7^{+11.2}_{-10.2}$) \kms\ (Table~\ref{tab:vmodel}), indicating that the wedge shape is roughly symmetric. That is, the inbound stars are moving at approximately the same speed as the outbound stars. 

In addition, the right panel of Figure~\ref{fig:wedge} highlights M31 RGB stars that probably belong to kinematical substructure (Section~\ref{sec:mcmc}) dominated by the NE shelf (Section~\ref{sec:phot_metal}). The upper envelope appears to have a higher density of stars than the lower envelope (see also Figure~\ref{fig:vmodel}), where the lower envelope only becomes statistically distinguishable from the background kinematically hot population at \Rproj\ $\sim$ 19 kpc in {\tt NE1}. 
However, the clustering of stars close to the expected velocity of the lower envelope in {\tt NE3} (\vhelio\ $\lesssim$ $-400$ \kms; Figure~\ref{fig:vmodel}) that have \fehphot\ consistent with the NE shelf (Figure~\ref{fig:vhelio_vs_fehphot}) implies that the lower envelope may be present at low levels in the data. 

\subsubsection{Constraints on the Gravitational Potential of M31}
\label{sec:wedge_model}

As first shown by \citet{MerrifieldKuijken1998}, hereafter MK98,
the observed line-of-sight velocities of stars in a tidal shell could in principle constrain the radial component of the gravitational potential ($g_s$) of the host galaxy at the shell radius ($r_s$). Thus, we utilized our measurements of M31-centric line-of-sight velocity ($v_{\rm los}$) for RGB stars in the NE shelf region as a test of such analytical models. 
The 
absolute maximum velocity of the wedge as a function of projected galactocentric radius $R$ is given by the following expression,

\begin{equation}
\label{eq:envelope}
v_{\rm los}^{\rm MK} = \sqrt{\frac{g_s}{r_s}} (r_s - R).
\end{equation}
MK98 derived Eq.~\ref{eq:envelope} assuming that a tidal shell is composed of stars on monoenergetic orbits in a plane tangential to the line of sight, where stars in the shell have low angular momentum such that the three-dimensional shell velocity ($v_s$) is zero. We adopted Eq.~\ref{eq:envelope} to describe the shape of the NE shelf wedge pattern instead of the formulation by \citet{SandersonHelmi2013}, hereafter SH13,

\begin{equation}
\label{eq:envelope_sh13}
v_{\rm los}^{\rm SH} = \frac{ \sqrt{r_s^2 + 2 R r_s - 3 R^2} (\sqrt{g(r_s) (r_s - R)} + v_s)}{r_s + R},
\end{equation}
which lifts the monoenergetic and low angular momentum assumptions, 
because the simpler MK model exhibits better convergence when applied to the data. The reason for this is likely that the tip of the wedge pattern, which places the strongest constraints on $v_s$, is not included in the NE shelf data (Section~\ref{sec:wedge_obs}), in conjunction with the fact that $v_s$ is instrinsically difficult to constrain with typical values of $\lesssim$ 30 \kms\ (SH13).


In order to estimate $g_s$, we therefore fit Eq.~\ref{eq:envelope} to the upper envelope of the NE shelf wedge pattern. We used the upper, as opposed to the lower, envelope owing to projection effects (SH13) that result in a minor asymmetry in which the positive edge has a greater maximum apparent speed for the innermost field (Figure~\ref{fig:wedge}), besides which the detection of the lower envelope in this field is uncertain (Section~\ref{sec:wedge_obs}). We used the kinematical parameters from Table~\ref{tab:vmodel} to define the upper envelope using the following equation for each NE shelf field
\begin{equation}
\label{eq:up_env}
v_{\rm los} = \mu_{v, {\rm KCC2}} + 2\sigma_{v, {\rm KCC2}} - v_{\rm M31},
\end{equation}
 where we calculated $\delta v_{\rm los}$ using standard error propagation by assuming symmetric average uncertainties on $\mu_{v, {\rm KCC2}}$ and $\sigma_{v, {\rm KCC2}}$. 
 We assessed the possibility of disk contamination in KCC2 of {\tt NE1} by conservatively removing all stars with \fehphot\ in the range expected for the majority of disk stars 
 (Section~\ref{sec:phot_metal}) and found that the upper envelope location was unchanged.

Thus, we used the non-linear least squares Python algorithm {\tt curve\_fit} \citep{scipy} to obtain $g_s =$ \gsup\ $\pm$ \gsuperr\ km s$^{-1}$ Myr$^{-1}$ and $r_s = $ \rsup\ $\pm$ \rsuperr\ kpc from the observed shape of the NE shelf wedge. The right panel of Figure~\ref{fig:wedge} shows the data points corresponding to the upper edge of the feature and its best-fit model. Assuming a spherical potential, we can transform these constraints into an upper limit on M31's enclosed mass ($M_{\rm enc}$) at the shell radius,

\begin{equation}
    M_{\rm enc}(r < r_s) = \frac{g_s r_s^2}{G},
\end{equation}
where we found that $M_{\rm enc}$($r <$ 36.6 kpc) = (\Menc\ $\pm$ \dMenc) $\times$ 10$^{11}$ $M_{\odot}$ based on the NE shelf data.

We note that these parameter estimates should be treated with caution given the assumptions involved in the derivation of Eq.~\ref{eq:envelope}. SH13 used the \citet{Fardal2007} GSS merger simulations to compare the value of $g_s$ derived from the projected phase space distributions of predicted caustics to the true values in the simulation. They found that the MK model overestimated the value of $g_s$ by a factor of $\sim$2-3, whereas the more sophisticated SH13 model produced more accurate estimates.
Thus, the empirically derived value of $g_s$ should be interpreted as an upper limit. We compare this value to 
literature measurements of M31's enclosed mass in Section~\ref{sec:dark_matter}.

\section{N-body Models for the Formation of the Northeast shelf}
\label{sec:nbody}

\begin{figure*}
    \centering
    \includegraphics[width=\textwidth]{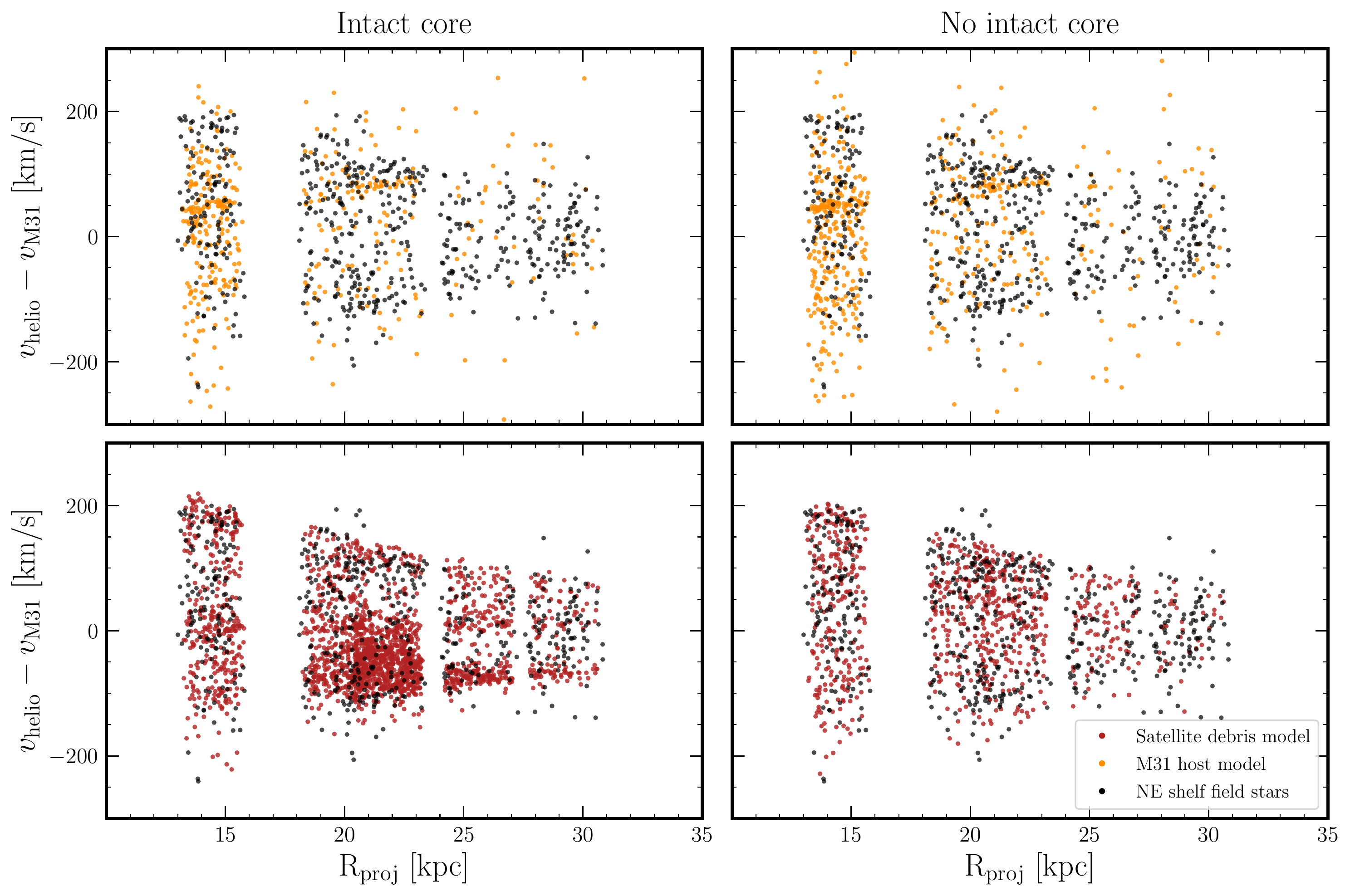}
    \caption{Similar to Figure~\ref{fig:nbody_wedge}, except we restricted the selection region for the NE Shelf in the N-body models 
    to the location of the spectroscopic fields (Figure~\ref{fig:fields}). Top panels (bottom panels) show the comparison to the M31 host (satellite debris) models with (left panels) and without (right panels) an intact core.
    The NE shelf region in the simulations is dominated by tidal debris, as opposed to the disk or phase-mixed stellar halo, at the data location, although the models predict a non-negligible disk contribution from $v_{\rm helio} - v_{\rm M31} \sim +(50 - 150)$ \kms\ at R$_{\rm proj}$ $\lesssim$ 23 kpc (Section~\ref{sec:nbody_wedge}).
    }
    \label{fig:nbody_wedge_deimosarea}
\end{figure*}

We compared observations of RGB stars in the NE shelf to predictions of N-body models for the GSS merger event to test scenarios for the shelf's formation. 
We utilized re-simulations of models from \citet{Fardal2007}, hereafter F07, and \citet{Fardal2012}, hereafter F12, which were constructed using a static bulge–disc–halo model for M31’s gravitational potential \citep{Geehan2006} and a spherical Plummer model for a satellite progenitor consisting of only a stellar component \citep{Fardal2006}. A particle realization for the M31 host component was incorporated following the conclusion of the dynamical evolution of the N-body satellite model. In the F07 (F12) model, complete (partial) disruption of a satellite with stellar mass $M_s = 2.2 \times 10^9\ M_\odot$ ($3.1 \times 10^9\ M_\odot$) occurs 0.84 Gyr (0.97 Gyr) ago within a M31 host with stellar mass $M_h = 1.1 \times 10^{11}\ M_\odot$ and virial mass $M_{200} = 1.7 \times 10^{12}\ M_\odot$ for each simulation.

For the F07 and F12 models, respectively, the orbit of the progenitor was fitted to reproduce the observed properties of the GSS, {\it not} those of the NE, W, or SE shelves. Despite this, the models have shown good agreement with kinematics of RGB stars in the SE and W shelves \citep{Gilbert2007,Fardal2012} and promising overlap with PNe kinematics for a small sample in the NE shelf region \citep{Merrett2003,Merrett2006,Fardal2007,Fardal2013}. Figure~\ref{fig:nbody_roadmap} shows the satellite debris distribution on the sky for the F07 model, where the distribution for the F12 model is similar.



\subsection{Predicted Projected Phase Space Distributions}
\label{sec:nbody_wedge}

\begin{figure*}
    \centering
    \includegraphics[width=\textwidth]{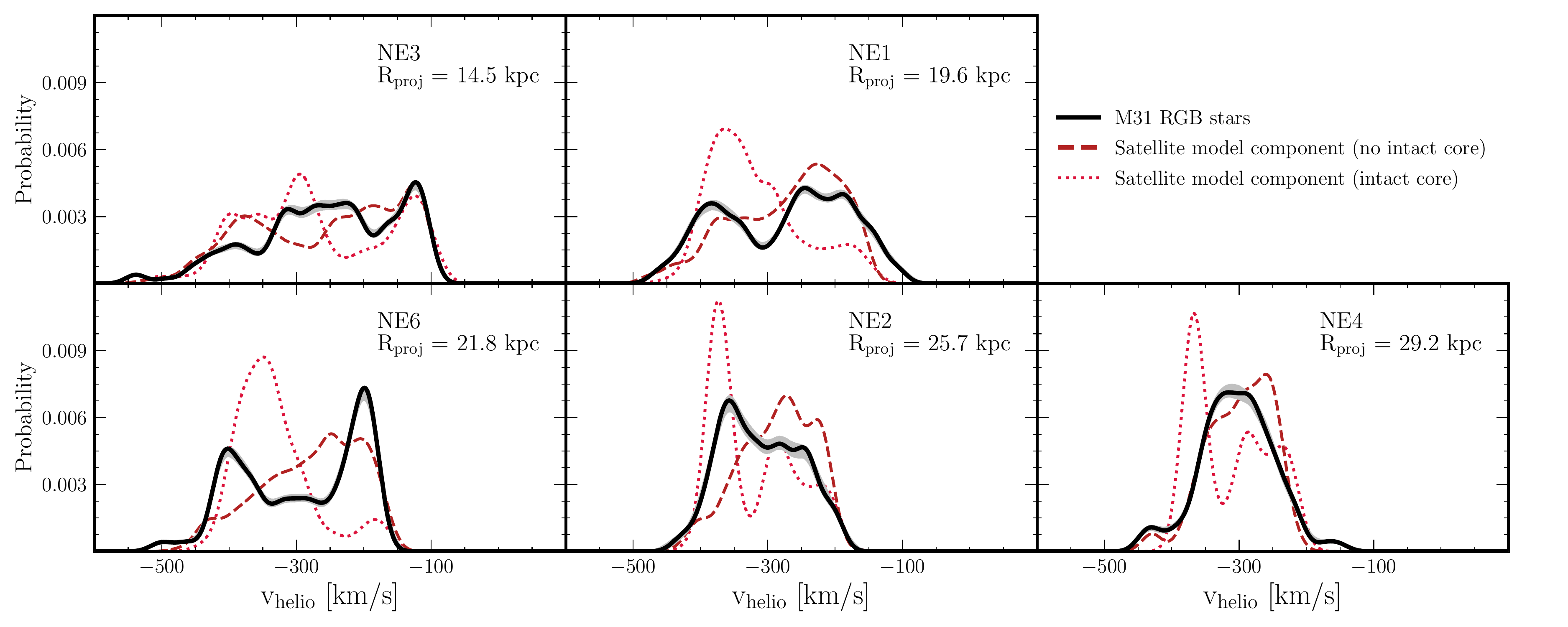}
    \caption{Line-of-sight velocity distributions predicted for 
    the NE shelf at the field locations (Figure~\ref{fig:nbody_wedge_deimosarea}) by N-body models for the formation of the GSS \citep{Fardal2007,Fardal2012} compared to observations of M31 RGB stars in the NE shelf (cf.\@ Figure~\ref{fig:vmodel}). We represent the velocity distributions using Gaussian kernel density estimation with $\sigma$ = 15 \kms.
    We show the predicted velocity distributions for the satellite debris components of the models with (red dotted lines) and without (red dashed lines) an intact core over the radial range spanned by each spectroscopic field. We also show the 1$\sigma$ confidence intervals (grey shaded regions) from the velocity measurement uncertainty on the observed distributions (solid black lines). 
    Neither model is able to reproduce the velocity distributions in detail, although the model without an intact core is a better match (Section~\ref{sec:nbody_core}).
    }
    \label{fig:nbody_vlos}
\end{figure*}

To assess whether this agreement extends to the kinematics of RGB stars in the NE shelf over a larger spatial extent, we compared the predicted projected phase space distributions to the observations in Figure~\ref{fig:nbody_wedge}. We also show the PNe observations from \citet{Merrett2003,Merrett2006} identified 
as a possible continuation of the GSS. We initially selected the NE shelf region in the simulations using the criteria $X_{\rm M31} >$ 0.2$^\circ$ and $-1.5^\circ < Y_{\rm M31} < 0.5^\circ$ (Figure~\ref{fig:nbody_roadmap}), where $X_{\rm M31}$ and $Y_{\rm M31}$ are defined along the NE major axis (P.A.\@ = 38$^\circ$ E of N) and NW minor axis in the tangent plane. We first chose to inspect the NE shelf over a larger spatial area than spanned by the spectroscopic fields for better statistical representations of the models. 
Additionally, we separated the N-body model particles into M31 host and GSS-related satellite debris components in Figure~\ref{fig:nbody_wedge}. 

For models with (F12) and without (F07) an intact satellite progenitor core, the projected phase space distribution of the tidal debris clearly provides a qualitative match to the NE shelf data. This not only supports the empirically motivated conclusion that the NE shelf is indeed a tidal shell (Section~\ref{sec:wedge_obs}), but also provides compelling evidence in favor of a GSS-related origin.
For the first time, we present a complete detection of the characteristic wedge pattern in the sense that it includes the returning stream component (i.e., the positive-velocity upper envelope) of the NE shelf. However, Figure~\ref{fig:nbody_wedge} supports the notion that the NE shelf observations do not resolve the tip of the wedge pattern (Section~\ref{sec:wedge_obs}) predicted to reside at R$_{\rm proj}$ $\gtrsim$ 35 kpc, where the tip is required to provide constraints on the 3D shell velocity (Section~\ref{sec:wedge_model}). Figure~\ref{fig:nbody_wedge} also corroborates the fact that the predicted negative-velocity caustic at R$_{\rm proj} \sim$ 15 kpc does not appear to be detected in the data (Section~\ref{sec:vel_model}). In general, the modeled upper envelopes of the wedge pattern appear to provide better fits to the data than the lower envelopes.

In order to explore predictions more equivalent to the observations,
we restricted the NE shelf selection region in the models to an area similar to the spatial coverage of the spectroscopic fields in Figure~\ref{fig:nbody_wedge_deimosarea}.
We found that the modeled disk contribution is substantially lower at the location of the NE shelf fields (in contrast to the broad selection region used in Figure~\ref{fig:nbody_wedge}). In the model with (without) an intact core, the host component constitutes \fhostcore\% (\fhostnocore\%) of the total particles, indicating that the expected satellite debris density is higher for the intact core model (Section~\ref{sec:nbody_core}). The M31 host model predicts a non-negligible disk component at $v_{\rm helio} -\ v_{\rm M31} \sim +30$, $+55$, and $+70$ \kms\ that overlaps with the primary, secondary, and secondary KCCs in {\tt NE3}, {\tt NE1}, and {\tt NE6}, respectively, all of which are suspected of being contaminated by the disk (Section~\ref{sec:disk}, Section~\ref{sec:phot_metal}). The contamination from the host decreases with projected radius, where the specific fractional contributions depend on the satellite model assumptions. Furthermore, the relative fraction of host versus satellite debris material is highly dependent on model assumptions regarding M31's mass and structural components.

\subsection{Remnant Core or Complete Disruption of the Satellite Progenitor?}
\label{sec:nbody_core}

 We evaluated whether the model with (F12) or without (F07) an intact satellite progenitor core provides a better fit to the observations. Figures~\ref{fig:nbody_wedge} and~\ref{fig:nbody_wedge_deimosarea} show that the primary difference between the projected phase space distributions of models with and without an intact core is a relative enhancement of structure (cf.\@ \citealt{Fardal2013}) at $v_{\rm helio} -\ v_{\rm M31} \lesssim 0$ \kms\ and an associated dearth of material approaching the tip of the wedge pattern. 
 From an initial inspection, the model without an intact core appears to provide a better match to the data, given the lack of obvious clumping in the interior of the observed wedge pattern (excepting perhaps the third component of unclear origin in {\tt NE3}; Section~\ref{sec:vel_model}). 

We further demonstrate the incompatibility of the intact core model in Figure~\ref{fig:nbody_vlos}, where we show the predicted versus observed line-of-sight velocity distributions for the NE shelf. The observed distribution is constructed from all RGB stars in the NE shelf fields, given evidence that the population is dominated by tidal debris (Section~\ref{sec:phot_metal}). The model distributions are constructed using only the satellite debris component at the data location (used in Figure~\ref{fig:nbody_wedge_deimosarea}).
In each case, we represent the velocity distributions using Gaussian kernel density estimation with $\sigma$ = 15 \kms. 
We calculated the uncertainty in the observed distribution via 10$^{3}$ perturbations by the (Gaussian) velocity uncertainties and re-evaluations of the kernel-smoothed probability density functions. The predicted line-of-sight velocity distributions constructed when including the host component are qualitatively consistent with those based solely on the satellite component for the F12 model. The main change for the F07 model is the addition of a peak at \vhelio\ $\sim$ $-270$ \kms\ in {\tt NE3} caused by the disk.

Although Figure~\ref{fig:nbody_vlos} shows that the models can roughly mimic the behavior of the observed line-of-sight velocity as a function of radius, particularly in terms of the width of the distributions, neither model provides an ideal match to the data. For example, the prominence and location of the negative-velocity edges of the modeled distributions, which correspond to the lower envelopes of the wedge patterns, do not approximate the data well. This disagreement is more pronounced for the F12 model than the F07 model. The F12 model also predicts more structure near M31's systemic velocity for \Rproj\ $\lesssim$ 22 kpc than is observed in the distributions.

We quantified the degree of similarity between modeled and observed velocity distributions as a function of radius using an Anderson-Darling test. We performed 10$^{3}$ trials in which we computed the test statistic between the perturbed observed velocity distributions and a random sampling of equal size from the models at the data location. 
The null hypothesis that the F12 model with an intact core is drawn from the same underlying distribution as the data can be rejected below the $\sim$5\% level in {\tt NE1}, {\tt NE6}, and {\tt NE4}, but not below the $\sim$15\% level for the other fields. The F07 model can be rejected at the $\sim$5\% level for {\tt NE2} but is consistent above the 10\% for all other fields. Including the host component does not alter these conclusions for either model.

\begin{figure*}
    \centering
    \includegraphics[width=\textwidth]{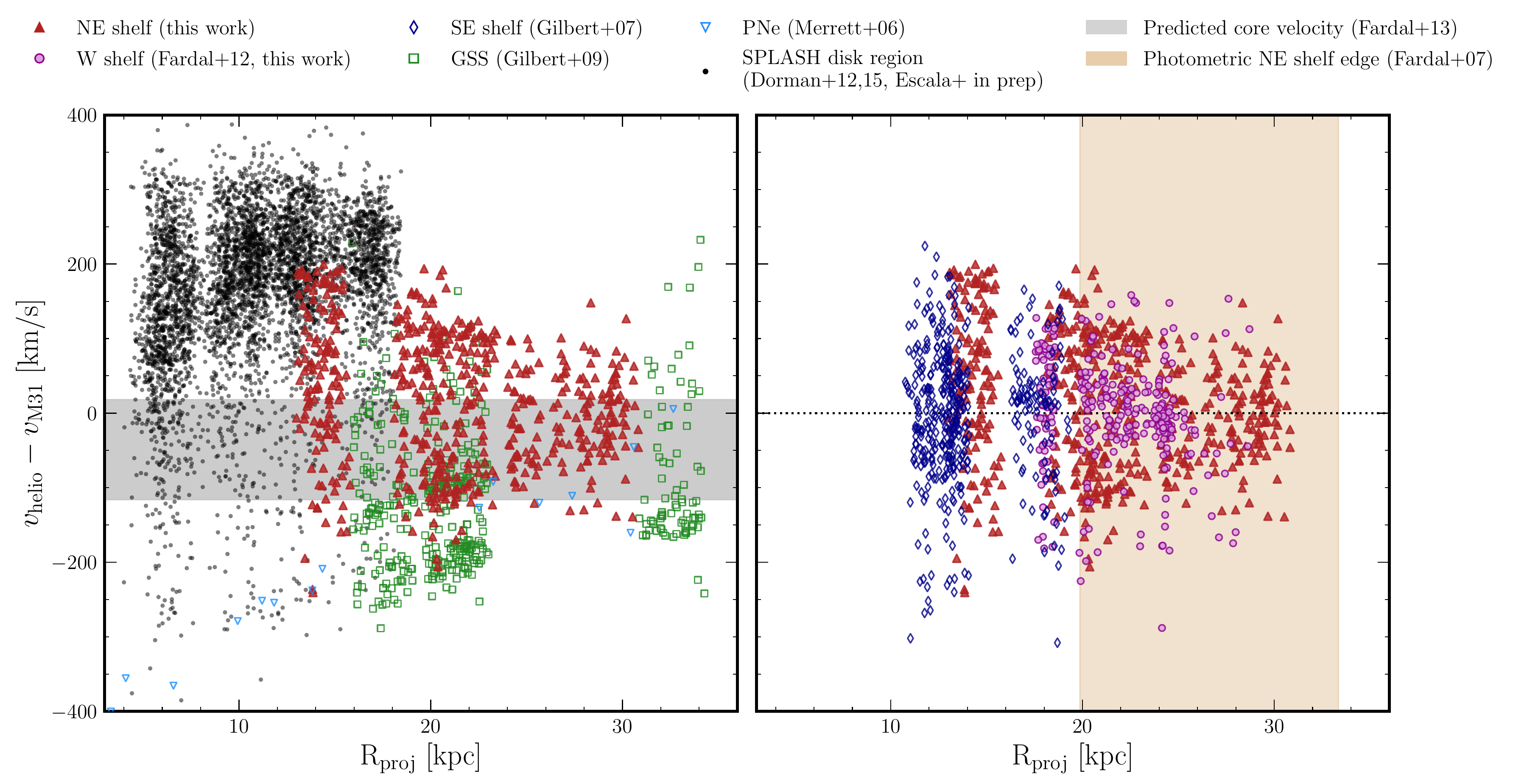}
        \caption{Heliocentric velocity relative to M31's systemic velocity versus projected M31-centric radius for M31 RGB stars (Section~\ref{sec:comp_phase_space}) along the line-of-sight to the GSS (green open squares; \citealt{Gilbert2009}), the NE (red filled triangles; this work), W (outlined magenta circles; F12), and SE (blue open diamonds) shelves, and the region of M31's northeastern disk surveyed by SPLASH (black points; \citealt{Dorman2012,Dorman2015}). We also show PNe that are possibly GSS-related (blue open inverted triangles; \citealt{Merrett2003,Merrett2006}). Each RGB sample contains stars belonging to the phase-mixed stellar halo in addition to substructure (i.e., the disk and/or tidal debris). (Left panel) The NE shelf compared to the GSS and the disk region. The secondary kinematical component along the GSS known as the KCC \citep{Kalirai2006,Gilbert2009} is located at $v_{\rm helio} - v_{\rm M31} > -150$ \kms\ and R$_{\rm proj} < 25$ kpc, overlapping with the lower envelope of the NE shelf. We do not find clear evidence for a continuation of the NE shelf in the disk region or an intact progenitor core based on its predicted velocity range (Section~\ref{sec:nbody_core}; \citealt{Fardal2013}). (Right panel) The shell patterns for the NE, W, and SE shelves, which are predicted to correspond to successive orbital wraps of GSS-related tidal debris. We show the projected radial range of the NE shelf edge inferred from imaging by F07 for reference.
    }
    \label{fig:comp_rproj_vs_vhelio}
\end{figure*}

In summary, the best available observational evidence favors a scenario in which the satellite progenitor that produced the NE shelf has completely disrupted, or at least is not present in the NE shelf region probed in this work, although comparison with a larger suite of N-body models is necessary for confirmation. If the intact core of the progenitor does exist, it could instead be embedded in the disk at $v_{\rm helio} - v_{\rm M31} \sim -49 \pm 67$ \kms\ compared to typical disk velocities of $v_{\rm helio} - v_{\rm M31} \sim +80 \pm 40$ \kms\ at predicted core positions \citep{Fardal2013}.
In the SPLASH survey of M31's disk, \citet{Dorman2012} tentatively detected NE shelf tidal debris as a forward continuation of the GSS at \vhelio\ $< -500$ \kms, but did not find evidence of a velocity signature resembling that of an intact core.
We further explore the possibility that the progenitor core is present in the SPLASH disk data in Section~\ref{sec:comp_phase_space}. We also discuss literature models for the NE shelf's formation in the context of our observations in Section~\ref{sec:nbody_lit}.

\section{Comparison to the Disk, West and Southeast Shelves, and the Giant Stellar Stream}
\label{sec:w_se_gss}

\begin{figure*}
    \centering
    \includegraphics[width=\textwidth]{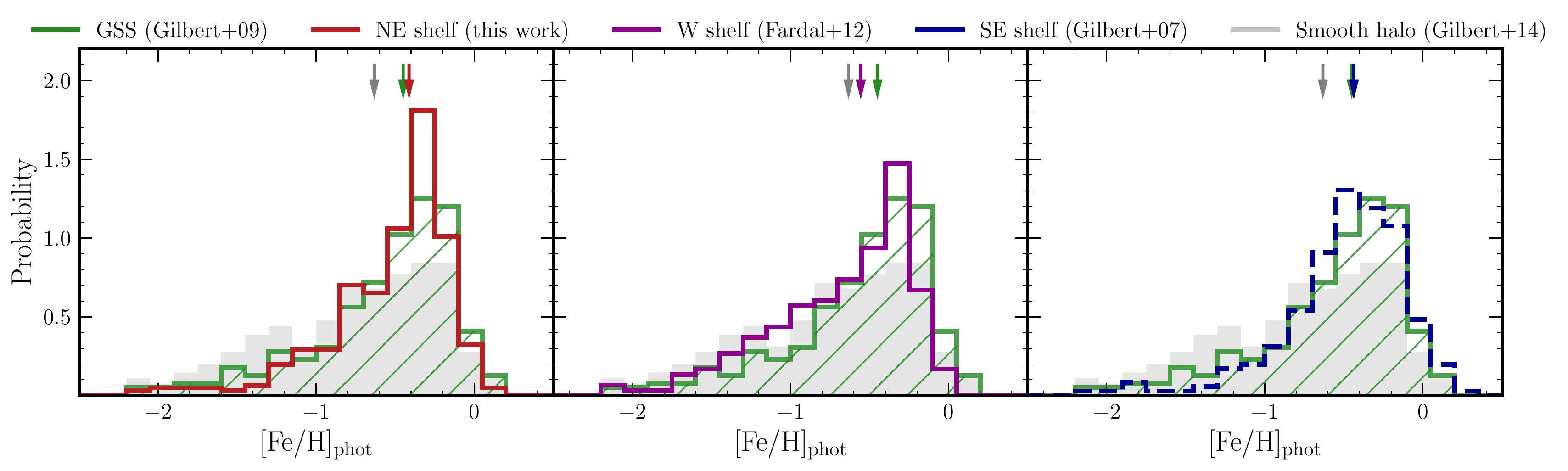}
    \caption{Photometric MDFs for the NE shelf (red histogram, left panel), W shelf (magenta histogram, middle panel), and SE shelf (blue histogram, right panel) compared to the GSS (green hatched histograms) and the smooth halo (gray filled histograms). Each stream/shelf sample consists of stars classified as M31 members (Section~\ref{sec:mem},~\ref{sec:comp_phase_space}) that probably belong to a given tidal feature ($p_{\rm sub} > 0.5$; Section~\ref{sec:mcmc},~\ref{sec:comp_phot}). The smooth halo sample is made from M31 members in SPLASH halo fields spanning 12--33 kpc \citep{Gilbert2012} with $p_{\rm sub} < 0.2$. We determined \fehphot\ for the smooth halo, GSS, W shelf, and SE shelf homogeneously with the NE shelf (Section~\ref{sec:phot}). The median \fehphot\ for each sample is shown as arrows (Table~\ref{tab:comp_phot}). The NE shelf has the same metallicity as the GSS and the SE shelf and is distinct from the smooth halo. The median \fehphot\ of the W shelf MDF is lower due to contamination by metal-poor halo stars, but shares a similar MDF peak.
    }
    \label{fig:comp_mdfs}
\end{figure*}

\subsection{Projected Phase Space Distributions}
\label{sec:comp_phase_space}


We compared the projected phase space distribution of the NE shelf to the W and SE shelves and the GSS to explore the relationship between the likely associated tidal features. Figure~\ref{fig:comp_rproj_vs_vhelio} shows M31-centric heliocentric velocity versus projected M31-centric radius for the NE shelf (this work), the GSS (\citealt{Gilbert2009}, hereafter G09), and the W (F12) and SE (\citealt{Gilbert2007}, hereafter G07) shelves, in addition to the stream PNe from \citet{Merrett2003,Merrett2006}. We also compared the NE shelf to M31's disk as surveyed by SPLASH \citep{Dorman2012,Dorman2015} given that the intact core of the GSS progenitor may be embedded in the disk (Section~\ref{sec:nbody_core}). 

For each feature, we show only stars that are more likely to belong to M31 than the MW foreground. For the GSS and SE shelves, M31 membership was evaluated using the likelihood-based method of \citet{Gilbert2006} as part of the SPLASH survey of M31's halo.\footnote{We constructed samples from spectroscopic fields (Figure~\ref{fig:fields}) {\tt f207}, {\tt H13s}, and {\tt a3} for the GSS \citep{Guhathakurta2006,Kalirai2006,Gilbert2009} and {\tt H11}, {\tt f116}, and {\tt f123} for the SE shelf \citep{Kalirai2006,Gilbert2007}. We classified stars with likelihoods $\langle L_i \rangle > 0$ when including radial velocity as a diagnostic \citep{Gilbert2006} as M31 members.} In contrast to F12, we utilized the Bayesian technique by \citet{Escala2020b} to identify M31 stars (Section~\ref{sec:mem}) in the W shelf,\footnote{Unlike the NE shelf (Section~\ref{sec:mem}), likelihood-based methods calibrated to the SE quadrant of M31's halo \citep{Gilbert2006,Escala2020b} can be reliably applied to the W shelf when including radial velocity as a diagnostic. The W shelf velocity distribution does not significantly overlap with the MW foreground and the stellar surface density (and expected MW contamination) of the halo in the NW quadrant is similar to the SE.} where this approach produces similar results to the original \citeauthor{Gilbert2006} method. We excluded the innermost W shelf field (Figure~\ref{fig:fields}) from our analysis owing to contamination by NGC 205 (F12). For stars in the SPLASH disk region, membership was determined from CMD position and the absorption strength of the Na I $\lambda$8190 doublet (I.~Escala et al., in preparation). We note that we do not distinguish between RGB stars belonging to substructure (i.e., the disk and/or tidal debris) and the phase-mixed halo in Figure~\ref{fig:comp_rproj_vs_vhelio}.

The right panel of Figure~\ref{fig:comp_rproj_vs_vhelio} suggests that the NE, W, and SE shelves form a sequence of tidal shells at successively lower energy. This fits with the prediction that they correspond to GSS progenitor tidal debris approaching its second, third, and fourth pericentric passage, respectively (e.g., F07, \citealt{Fardal2013}). The NE shelf is the outermost shell, extending to R$_{\rm proj} \gtrsim 31$ kpc and spanning $\lesssim$ 400 \kms, whereas the W and SE shelves are contained within R$_{\rm proj} \sim 25$ kpc and $\sim 18$ kpc respectively and $|v_{\rm helio} - v_{\rm M31}| \sim 100$ \kms. For reference, we show the projected radial range of the NE shelf edge inferred from imaging by F07, where the geometrical orientation of the shell relative to the line-of-sight causes this apparent variation in  R$_{\rm proj}$. Regardless of whether GSS formation models can match detailed observations in the shelves  (Section~\ref{sec:nbody}), their concentric wedge patterns provide compelling evidence for a shared origin in the tidal disruption of a single satellite galaxy.

The left panel of Figure~\ref{fig:comp_rproj_vs_vhelio} shows the NE shelf compared to the GSS and the disk region. The GSS fields include a secondary kinematical component known as the KCC that is likely physically associated with the GSS based on their tight correlation in \Rproj\ versus \vhelio\ space (Figure~\ref{fig:comp_rproj_vs_vhelio}) and chemical similarities
\citep{Kalirai2006,Gilbert2009,Gilbert2019}. Additionally, the potential forward continuation of the GSS speculated to be the NE shelf by \citet{Dorman2012} (Section~\ref{sec:nbody_core}) is visible in the disk data at $v_{\rm helio} - v_{\rm M31} < -200$ \kms\ and \Rproj\ $\lesssim$ 15 kpc. However, this group of stars does not seem to be consistent with an inward extension of the lower envelope of the NE shelf wedge pattern as probed by our data. The lower envelope of the NE shelf instead overlaps in phase space with the KCC. 

In a minor merger scenario for the formation of the GSS, the KCC could originate from an extension of the shelves (G09). The W shelf is the most promising candidate because its extent is not well-constrained owing to obscuration by M31's southwestern disk. Moreover, GSS formation models predict negligible contributions from the NE and SE shelves along the GSS (F07, F12). For the KCC to originate from a symmetric extension of the NE shelf, it would minimally require that the positive velocity caustic of the shell was depopulated at the KCC location.
Additional GSS formation models will be necessary to evaluate this possible origin for the KCC.

Figure~\ref{fig:comp_rproj_vs_vhelio} also shows the predicted velocity range of an intact GSS progenitor core, $v_{\rm helio} - v_{\rm M31} = -49 \pm 67$ \kms\ \citep{Fardal2013} compared to the expanded SPLASH disk dataset (i.e., including spectroscopic fields published by \citealt{Dorman2015}). As in the case of the NE shelf (Section~\ref{sec:nbody_core}), we do not find evidence for the presence of an intact core in the SPLASH data based on current theoretical predictions. The apparent clumps of stars at \Rproj\ $\sim$ 5--8 kpc within the predicted velocity range are probably associated with the disk, which approaches more halo-like velocities in this radial range \citep{Dorman2012}. Such clumps in velocity space also do not correspond to any spatial concentrations that would resemble a core. Moreover, the most likely locations for the core position are nearer to the NE shelf, not within the inner 5--8 kpc of the disk, although this radial range is not excluded by the models \citep{Fardal2013}. We note that \citet{Davidge2012} identified an overdensity at \Rproj\ $\sim 3.5$ kpc that is a plausible candidate for the GSS core, but this is outside the innermost extent of the SPLASH survey region.

\subsection{Photometric Metallicity Distributions}
\label{sec:comp_phot}

\begin{table}
 \centering
 \caption{Photometric Metallicity Distribution Properties for the Smooth Halo, GSS and NE, W, and SE Shelves}
\begin{threeparttable}
    \begin{tabular*}{\columnwidth}{lcccc}
    \hline \hline
    Feature & \fehphot$_{\rm,med}$ & $\langle$\fehphot$\rangle$ & $\sigma_{\rm [Fe/H]_{phot}}$\\
    \hline
    Smooth Halo & $-0.63^{+0.03}_{-0.05}$ & $-0.75 \pm 0.03$ & $0.53 \pm 0.02 $\\
    GSS & $-$0.43$^{+0.02}_{-0.03}$ & $-$0.56 $\pm$ 0.02 & 0.45 $\pm$ 0.02 \\
    NE shelf & $-$0.42 $\pm$ 0.01 & $-$0.53 $\pm$ 0.02 & 0.39 $\pm$ 0.02 \\
    W shelf & $-$0.55 $\pm$ 0.04 & $-$0.67$^{+0.02}_{-0.03}$ & 0.43 $\pm$ 0.02\\
    SE shelf & $-$0.44$^{+0.02}_{-0.01}$ & $-$0.50 $\pm$ 0.02 & 0.39 $\pm$ 0.02 \\
    \hline \hline
    \end{tabular*}
    \begin{tablenotes}[flushleft]
    \footnotesize
    \item Note.\@ \textemdash\ Similar to Table~\ref{tab:phot}, except comparing the NE shelf to likely associated tidal features and the smooth halo. We measured \fehphot\ homogeneously for each feature following Section~\ref{sec:phot}. 
    References for the original data: Halo \citep{Gilbert2014}, GSS \citep{Gilbert2009}, W shelf \citep{Fardal2012}, SE shelf \citep{Gilbert2007}.
    \end{tablenotes}
    \label{tab:comp_phot}
\end{threeparttable}
\end{table}

We assessed whether the \fehphot\ distributions of M31's giant stream and shell system support the prediction that each feature originates from the disruption of a single progenitor. Figure~\ref{fig:comp_mdfs} shows MDFs for the NE, W, and SE shelves compared to the GSS and the phase-mixed (``smooth'') component of the stellar halo. We assembled the smooth halo sample from M31 members in SPLASH halo fields spanning 12--33 kpc \citep{Gilbert2012,Gilbert2014}.\footnote{This radial range spans the following spectroscopic fields: {\tt H11}, {\tt f115}, {\tt f116}, {\tt f207}, {\tt f135}, {\tt H13s}, {\tt f130}, {\tt a0}, and {\tt a3}.} We measured \fehphot\ homogeneously for each sample following Section~\ref{sec:phot}. In order to 
separate tidal debris from smooth halo stars, we calculated substructure probabilities ($p_{\rm sub}$; e.g., Section~\ref{sec:mcmc}) for all RGB stars (Section~\ref{sec:comp_phase_space}) along the line-of-sight to each sample. For the GSS, SE shelf, and smooth halo, we employed G18's velocity distribution models to determine $p_{\rm sub}$, whereas we performed our own modeling of the velocity distribution for the W shelf (Appendix~\ref{appendix:wshelf_vmodel}). 

We defined the tidal debris MDFs using stars that are more likely to belong to substructure than to the kinematically hot halo-like population ($p_{\rm sub} > 0.5$). We adopted this more inclusive threshold for substructure membership (as opposed to $p_{\rm sub} > 0.75$) to maximize the sample size for the SE shelf, where its velocity signature at the base of its wedge pattern (\vhelio\ $\sim$ $-300$ \kms, $\sigma_v \sim 55$ \kms; G07, G18) makes it difficult to disentangle from the halo. The SE and W shelf MDFs are the most likely to suffer from halo contamination, where the maximum values of $p_{\rm sub}$ are $\sim$\psubmaxse\ and $\sim$\psubmaxw, respectively, whereas the GSS and NE shelf achieve maximum values of $\sim$\psubmaxgssne. We used $p_{\rm sub} < 0.2$ to conservatively define the smooth halo MDF.


Figure~\ref{fig:comp_mdfs} demonstrates that the peak metallicities (\fehphot\ $\sim$ $-0.40$; Table~\ref{tab:comp_phot}) and shapes of the MDFs are similar between the shelves and the GSS, whereas they are distinct from the smooth halo MDF (\fehphot\ $\sim$ $-0.6$). The median (mean) \fehphot\ of the W shelf is more metal-poor than the GSS and the other shelves, but this discrepancy likely results from increased contamination by comparatively metal-poor halo stars at \fehphot\ $\lesssim$ $-0.7$. Based on an iterative analysis in which substructure probabilities were calculated using both chemical and kinematical information, F12 amplified differences between the tidal debris and phase-mixed halo populations in the W shelf region to show that the \fehphot\ $\gtrsim$ $-0.7$ metallicity range is dominated by the W shelf. Previous analyses of the \fehphot\ distributions of the W and SE shelves (F12, G07) found agreement with the GSS, supporting a common origin scenario for these tidal features. \citet{Escala2021} also found agreement between spectral-synthesis based measurements of [Fe/H] and [$\alpha$/Fe] in the SE shelf and GSS as part of the Elemental Abundances in M31 survey.

Figure~\ref{fig:comp_mdfs} further shows that the metallicity distribution of the NE shelf broadly agrees with the GSS, and by extension the W and SE shelves. We tested the hypothesis that the NE shelf and GSS MDFs share a common origin using a $p_{\rm sub}$-weighted Anderson-Darling test with 10$^{3}$ bootstrap resamplings, finding that it could not be rejected below a 5\% significance level. This conclusion is not affected by the selection criterion for NE shelf and GSS stars (i.e., by halo contamination) given that these regions are heavily polluted by tidal debris (Section~\ref{sec:phot_metal}; G09). Thus, the chemical similarity between the NE shelf and the GSS/other shelves provides evidence in favor of a direct physical association between the tidal features,
especially given the independent corroboration by the projected phase space distributions (Section~\ref{sec:comp_phase_space}).

\section{Discussion}
\label{sec:discuss}

\subsection{The Enclosed Mass of M31}
\label{sec:dark_matter}

We compared the estimate of M31's enclosed mass from NE shell kinematics (Section~\ref{sec:wedge_model}), $M_{\rm enc} = $ (\Menc\ $\pm$ \dMenc) $\times 10^{11} M_{\rm \odot}$ at a shell radius $r_{\rm s}$ = \rsup\ $\pm$ \rsuperr\ kpc, to recent literature measurements of M31's virial mass (see e.g., \citealt{Evans2000} and \citealt{Klypin2002} for earlier measurements) from various kinematic tracers. From M31's satellite galaxies, \citealt{Watkins2010}, \citet{Tollerud2012}, and \citet{Patel2017} measured consistent virial masses in the 1$\sigma$ range of $M_{\rm 200} =$ (\Mvirsatlo---\Mvirsathi) $\times 10^{12} M_{\rm \odot}$. An analysis using globular cluster kinematics by \citet{Veljanoski2013} found agreement with values of $M_{\rm 200} = (1.0-1.7) \times 10^{12} M_{\rm \odot}$. Additionally, applying the timing argument to measure the masses of the MW and M31 simultaneously yields $M_{\rm 200,M31} = (1.5 \pm 0.4) \times 10^{12} M_{\rm \odot}$ when combined with previous observational estimates of M31's virial mass \citep{vdMG08,vdM12}. The Bayesian sampling approach of \citet{Fardal2013} to infer M31's virial mass from GSS merger models 
is also in accord with other measurements, where they found $\log_{\rm 10} (M_{\rm 200} / M_{\rm \odot}) =$ \mfardal\ $\pm$ \mfardalerr.

Owing to the general agreement between measurements of M31's virial mass and the fact that \citet{Fardal2013} similarly used tidal debris to obtain their measurement, we focused on comparisons to their analysis to place our mass estimate in context. We converted \citeauthor{Fardal2013}'s value into an estimate of the enclosed halo mass at the NE shell radius according to their assumed Navarro-Frenk-White (NFW) profile, finding $M_{\rm enc,halo}(r = r_s) =$ (\Mencfardal\ $\pm$ \Mencfardalerr) $\times 10^{11} M_{\rm \odot}$. To obtain the total enclosed mass including baryons, we incorporated the mass of the bulge associated with the NFW profile ($M_{\rm b} = 3.40 \times 10^{10} M_{\rm \odot}$) and the mass of the disk at the NE shell radius projected onto the disk plane. Given the total mass of the disk in the simulations ($M_{\rm d,tot} = 7.34 \times 10^{10} M_{\rm \odot}$), its scale length ($R_{\rm d} = 5.40$ kpc), and assuming $R_s$ = 40 kpc (the maximum value of \Rdisk\ in the field at the tip of the wedge pattern, {\tt NE4}), $M_{\rm d}(r = r_s) =$ \Mdiskatshell\ $\times 10^{10} M_\odot$. The enclosed mass at the NE shell radius deduced from \citealt{Fardal2013} is therefore $M_{\rm enc}(r = r_s) =$ (\Mencfardalbary\ $\pm$ \Mencfardalerr) $\times 10^{11} M_{\rm \odot}$. 

This calculation validates the interpretation of the MK-based approximation (Eq.~\ref{eq:envelope}) of the enclosed mass  as a likely overestimate (Section~\ref{sec:wedge_model}; F07, SH13) by a factor of \Mencoverestimate\ $\pm$ \dMencoverestimate. Future work could simultaneously leverage measurements from the NE, W, and SE shelves---either in combination with N-body models or more sophisticated numerical implementations of the shell equations---to improve constraints on M31's enclosed mass from shell kinematics.

\subsection{Implications for GSS Merger Scenarios}
\label{sec:nbody_lit}

We discuss the implications of the projected phase space and metallicity distributions of the NE shelf for aspects of GSS merger scenarios not considered in Sections~\ref{sec:nbody} and~\ref{sec:w_se_gss}. As previously established, the NE shelf metallicity distribution can place unique constraints on GSS formation models that track metallicity because the metal-rich central debris of the progenitor most likely pollutes this region \citep{Fardal2008,Miki2016,Kirihara2017}. For example, \citet{Miki2016} predicted that an observed negative metallicity gradient in the GSS \citep{Ibata2007,Gilbert2009,Conn2016,Cohen2018,Escala2021} should result in higher metallicity for the NE shelf than other GSS-related tidal structures, whereas the W shelf and GSS metallicity should be comparable (as found by F12).\footnote{The metallicity here refers to the high surface brightness regions of the GSS (which are used in this work; Figure~\ref{fig:fields}, Section~\ref{sec:comp_phase_space}), as opposed to its more metal-poor envelope.} In contrast, the NE shelf metallicity distribution appears to be consistent with that of the W shelf and GSS (Section~\ref{sec:comp_phot}) rather than being more metal-rich. However, more detailed explorations of the NE shelf MDF (Section~\ref{sec:phot_metal}) may reveal subtle variations on the sky that could be mapped to model predictions. 

Regarding the projected phase space distribution of the NE shelf, 
\citet{Fardal2013} found that N-body models with high values of satellite progenitor mass 
and central density corresponded to the presence of a large intact core in the NE shelf region. The lack of an apparent core in the NE shelf (Section~\ref{sec:nbody_core}) or the neighboring disk (Section~\ref{sec:comp_phase_space}) therefore supports a lower central density and mass for the progenitor such that it completely disrupts in the case of a minor merger. In a major merger, the progenitor core has been suggested to be fully disrupted \citep{Hammer2018} or intact in the form of M32 \citep{DSouzaBell2018}, although this latter possibility is in direct contradiction to the positions and velocities of central orbits consistent with the shelves in the \citeauthor{Fardal2013} models.

Both minor and major merger models have found that observational constraints from the GSS (and the disk for major mergers) allow for more variation in the NE shelf location than other shelves  \citep{Fardal2013,Hammer2018}. In the \citeauthor{Hammer2018} simulations, the NE shelf often appears smaller on the sky than observed, highlighting the need for major merger models to explore whether they can reproduce M31's shell system in detail. In any given scenario, the impact of significant internal rotation in the progenitor on the distribution of tidal debris in the shelves similarly warrants further examination (e.g., \citealt{Fardal2008,Kirihara2017,DSouzaBell2018}), where in this work we have only performed comparisons to models with spheroidal progenitors (Section~\ref{sec:nbody}). Thus, the combination of new observational constraints for the NE shelf presented in this work and existing constraints from prior studies of the W and SE shelves (G07, F12) have great potential for refining future GSS merger models.

\section{Summary}
\label{sec:summary}

We have measured radial velocities and photometric metallicities from Keck/DEIMOS spectra of \Ntot\ RGB stars spanning 13---31 projected kpc along the line-of-sight to M31's NE shelf. We have performed the first detailed kinematical and chemical characterization of RGB stars in the NE shelf and have presented the first complete detection of a ``wedge'' pattern in projected phase space (i.e., that includes the returning stream component of the shelf; Section~\ref{sec:wedge_obs}). This study presents conclusive evidence that (1) the NE shelf is indeed a tidal shell as inferred from prior studies and (2) the NE shelf forms a multiple shell system together with the W and SE shelves (Section~\ref{sec:comp_phase_space}). We have also found that:

\begin{enumerate}
    \item The photometric MDF of RGB stars along the line-of-sight to the NE shelf (median \fehphot\ = \fehphotmednepsub\ $\pm$ \fehphoterrmednepsub) is likely dominated by tidal debris as opposed to M31's disk or phase-mixed component of M31's stellar halo, although there is evidence for a low level of disk contamination (Section~\ref{sec:phot_metal}). 
    \item The MDF of the NE shelf is consistent with the GSS and W and SE shelves, supporting a direct physical association between the tidal features (Section~\ref{sec:comp_phot}).
    \item The projected phase space distribution of the NE shelf exhibits broad agreement with GSS formation models in which the NE shelf is the second orbital wrap of the progenitor (F07, F12; Section~\ref{sec:nbody_wedge}), thereby providing support for this origin in a minor merger scenario.
    \item There is currently no evidence for a kinematical signature in the NE shelf region (Section~\ref{sec:nbody_core}) or in the neighboring disk (Section~\ref{sec:comp_phase_space}) corresponding to an intact GSS progenitor core, thereby favoring formation models in which the progenitor is completely disrupted (Section~\ref{sec:nbody_lit}).
    \item Regardless of the degree to which the model predictions and observations of the NE shelf agree (Section~\ref{sec:nbody}), the simplicity of a common origin scenario for explaining the concentric wedge patterns of the NE, W, and SE shelves independently provides proof in its favor (Section~\ref{sec:comp_phase_space}).
    \item The projected phase space distribution of the NE shelf in combination with analytical models for tidal shell formation constrains M31's enclosed mass to an upper limit of (\Menc\ $\pm$ \dMenc) $\times$ $10^{11} M_{\odot}$ at a shell radius of \rsup\ $\pm$ \rsuperr\ kpc (Section~\ref{sec:wedge_model}), which is likely an overestimate by approximately a factor of two in agreement with expectations from simulations (Section~\ref{sec:dark_matter}).
\end{enumerate}

\begin{acknowledgments}
IE\@ acknowledges generous support from a Carnegie-Princeton Fellowship through the Carnegie Observatories. This material is based upon work supported by the NSF under Grants No.\ AST-1909759 (PG), AST-1909066 (KMG), and
AST-2007232 (RES). RES additionally acknowledges support from NASA grant 19-ATP19-0068, the Research Corporation through the Scialog Fellows program on Time Domain Astronomy, and from HST-AR-15809 through STScI.

We are grateful to the many people who have worked to make the Keck Telescope and its instruments a reality and to operate and maintain the Keck Observatory. The authors wish to recognize and acknowledge the very significant cultural role and reverence that the summit of Maunakea has always had within the indigenous Hawaiian community.  We are most fortunate to have the opportunity to conduct observations from this mountain.
\end{acknowledgments}

\vspace{5mm}
\facilities{Keck (DEIMOS), CFHT (MegaCam)}


\software{astropy \citep{astropy13,astropy18},  
          scikit-learn \citep{sklearn},
          emcee \citep{Foreman-Mackey2013},
          scipy \citep{scipy},
          numpy,
          matplotlib}

\appendix

\section{Empirical Modeling of the W Shelf Velocity Distribution}
\label{appendix:wshelf_vmodel}

\begin{figure}
    \centering
    \includegraphics[width=\textwidth]{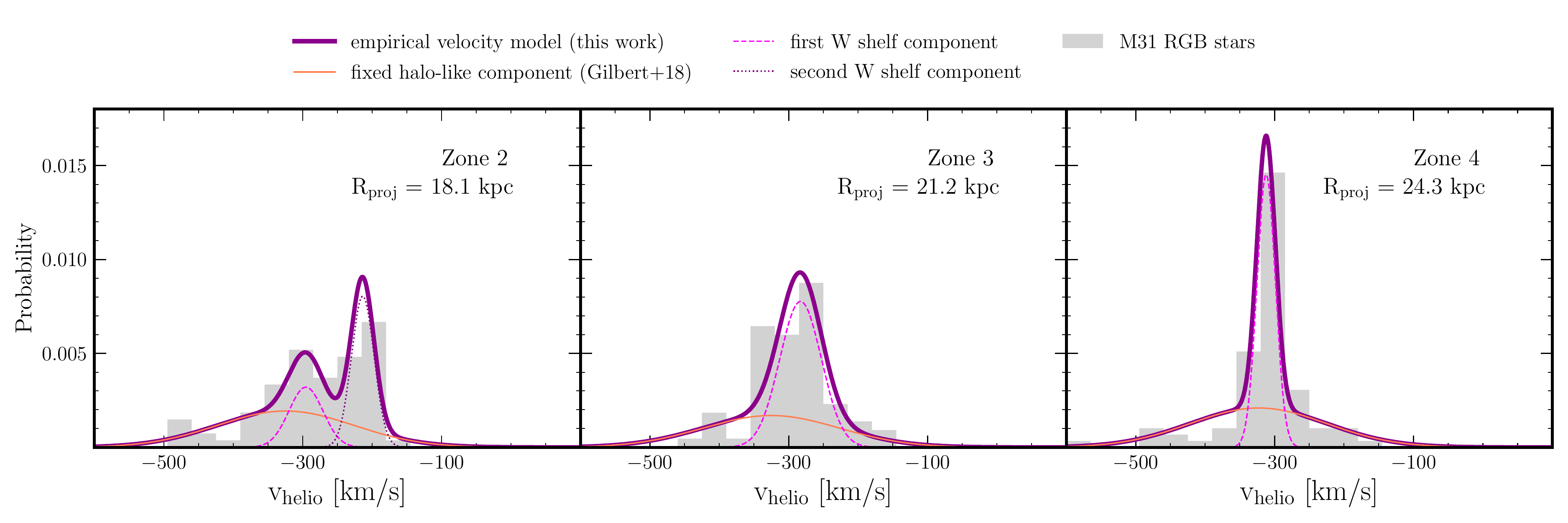}
    \caption{Same as Figure~\ref{fig:vmodel}, except for the W shelf. We excluded the innermost spectroscopic field targeting the W shelf (Figure~\ref{fig:fields}) from the analysis in Section~\ref{sec:w_se_gss} owing to contamination from NGC 205 (F12). We separated the W shelf fields into radial zones similar to those utilized by F12 for the velocity distribution modeling (Section~\ref{sec:vel_model}).
    }
    \label{fig:wshelf_vmodel}
\end{figure}

\begin{table*}
    \centering
    \caption{Empirical Velocity Distribution Model Parameters for the W Shelf}
    \begin{threeparttable}
    \begin{tabular*}{\textwidth}{@{\extracolsep{\fill}}lccccccccc}
    \hline\hline
    Field & \multicolumn{1}{p{1.0cm}}{\centering R$_{\rm proj}$\\(kpc)} & \multicolumn{1}{p{1.0cm}}{\centering $\mu_{\rm halo}$\\(km/s)} & \multicolumn{1}{p{1.0cm}}{\centering $\sigma_{\rm halo}$\\(km/s)} & \multicolumn{1}{p{1.0cm}}{\centering $\mu_{\rm KCC1}$\\(km/s)} & \multicolumn{1}{p{1.0cm}}{\centering $\sigma_{\rm KCC1}$\\(km/s)} & \multicolumn{1}{p{1.0cm}}{\centering $f_{\rm KCC1}$\\(km/s)} & \multicolumn{1}{p{1.0cm}}{\centering $\mu_{\rm KCC2}$\\(km/s)} & \multicolumn{1}{p{1.0cm}}{\centering $\sigma_{\rm KCC2}$\\(km/s)} & \multicolumn{1}{p{1.0cm}}{\centering $f_{\rm KCC2}$\\(km/s)}\\ \hline
    Zone 2 & 18.1 & $-$324.8 & 98.1 & $-$295.5$^{+26.8}_{-16.9}$ & 24.4$^{+14.0}_{-9.4}$ & 0.20$^{+0.13}_{-0.10}$ & $-$213.6$^{+4.9}_{-5.4}$ & 16.3$^{+5.1}_{-3.8}$ & 0.33$^{+0.09}_{-0.09}$\\
    Zone 3 & 21.2 & $-$325.3 & 98.1 & $-$282.9$^{+4.9}_{-4.7}$ & 30.0$^{+4.5}_{-4.0}$ & 0.58$^{+0.08}_{-0.08}$ & ... & ... & ...\\
    Zone 4 & 24.3 & $-$323.3 & 98.0 & $-$312.2$^{+2.4}_{-2.4}$ & 13.3$^{+2.7}_{-2.1}$ & 0.48$^{+0.07}_{-0.07}$ & ... & ... & ...\\
    \hline
    \end{tabular*}
    \begin{tablenotes}[flushleft]
    \footnotesize
    \item Note. \textemdash\ Same as Table~\ref{tab:vmodel}, except for the W shelf.
    \end{tablenotes}
    \end{threeparttable}
    \label{tab:wshelf_vmodel}
\end{table*}


In Section~\ref{sec:comp_phot}, we compared the photometric MDFs of the NE shelf to those of the GSS, SE shelf, and W shelf. This comparison requires the identification of RGB stars that probably belong to substructure in order to select a relatively clean sample of stars associated with the various tidal debris features. Although the observed velocity distributions of the GSS and SE shelf have been previously modeled empirically \citep{Gilbert2018} similar to the NE shelf (Section~\ref{sec:vel_model}), this is not the case for the W shelf. In order to calculate a MDF for the W shelf, F12 instead used their N-body model for the GSS merger to evaluate the likelihood that a RGB star belonged to GSS tidal debris versus their M31 host model constructed from a disk and halo component (Section~\ref{sec:nbody}). Thus, we empirically modeled the velocity distribution of the W shelf using a Gaussian mixture following Section~\ref{sec:vel_model}.

We separated the W shelf fields into radial zones analogous to those used by F12, where we discarded the innermost field (F12's Zone 1) from the analysis owing to contamination from NGC 205. We incorporated the outermost field (F12's Zone 5) into our Zone 4 given that this field primarily contains phase-mixed halo stars and therefore does not affect the mean and dispersion of the W shelf component over this radial range. The net effect of constructing the Zone 4 model additionally using stars in Zone 5 is to decrease $f_{\rm KCC}$ and slightly decrease $p_{\rm sub}$ for stars in both zones. This effect is negligible in practice owing to the $p_{\rm sub}$ threshold adopted in Section~\ref{sec:comp_phot}.

We assumed N$_{\rm comp}$ = (2, 1, 1) substructure components for Zones (2, 3, 4) in addition to a kinematically hot halo-like component adopted from \citet{Gilbert2018} for each zone (Section~\ref{sec:ncomp}). We note that the contribution of M31's disk to the velocity distribution of the W shelf fields is expected to be negligible (F12). 
Table~\ref{tab:wshelf_vmodel} presents the empirical velocity distribution model parameters for each radial zone and Figure~\ref{fig:wshelf_vmodel} shows the observed velocity distributions of each radial zone compared to its model. Given these models, we computed the probability that each RGB star in the W shelf fields belongs to substructure (Section~\ref{sec:mcmc}) for use in determining the MDF of the W shelf.


\end{document}